\documentclass[useAMS,usenatbib]{mn2e}
\usepackage{amsmath}
\usepackage{times}
\usepackage{psfig}
\usepackage{graphics,graphicx}
\usepackage{lscape}
\usepackage{epsf}
\usepackage{epsfig}
\usepackage{color}
\usepackage{dcolumn}
\usepackage{natbib}
\bibpunct{(}{)}{;}{a}{}{,}

\newcommand{\mdot}{\mbox{$\dot{M}$}}
\newcommand{\Rstar}{\mbox{$R_\ast$}}

\newcommand{\vinf}{\mbox{$v_\infty$}}
\newcommand{\xmm}{{\em XMM-Newton}}

\newcommand{\xicma}{\mbox{$\xi^1$\,CMa}}
\newcommand{\zcas}{\mbox{$\zeta$\,Cas}}
\newcommand{\voph}{\mbox{V2052\,Oph}}
\newcommand{\bcep}{\mbox{$\beta$\,Cep}}
\newcommand{\tsco}{\mbox{$\tau$\,Sco}}
\newcommand{\Teff}{\mbox{$T_{\rm eff}$}}
\newcommand{\Lbol}{\mbox{$L_{\rm bol}$}}

\newcommand{\Lx}{\mbox{$L_{\rm X}$}}

\newcommand{\Msun}{\mbox{$M_\odot$}}
\newcommand{\myr}{\mbox{$M_\odot\,{\rm yr}^{-1}$}}
\newcommand{\lsim}{\raisebox{-.4ex}{$\stackrel{<}{\scriptstyle \sim}$}}
\newcommand{\msim}{\raisebox{-.4ex}{$\stackrel{>}{\scriptstyle \sim}$}}

\newcommand{\flux}{erg\,cm$^{-2}$\,s$^{-1}$}
\newcommand{\captionabove}[1]{\caption{#1}}

\def \etal   {\hbox{et~al.\/}}
\def\changed{}

\title[Early magnetic B-type stars]{Early magnetic B-type stars:
X-ray emission and wind properties}
\author[L. M. Oskinova \etal\ ]
{L. M. Oskinova$^{1}$,\thanks{E-mail: lida@astro.physik.uni-potsdam.de}
H. Todt$^{1}$,  R.~Ignace$^{2}$, J.\,C.\,Brown$^{3}$, J.\,P.\, Cassinelli$^{4}$,
W.-R.\,Hamann$^{1}$  \\
$^{1}$ Institute for Physics and Astronomy, University of Potsdam,
14476 Potsdam, Germany \\
$^{2}$ Department of Physics and Astronomy, East
Tennessee State University,
Johnson City, TN 37614, USA\\
$^{3}$ School of Physics and Astronomy, University of Glasgow, Glasgow G12 8QQ, UK\\
$^{4}$ Department of Astronomy, University of Wisconsin-Madison, Madison, WI 53711, USA}

\begin{document}

\date{Accepted . Received ; in original form 21.02.2011 22:32}

\pagerange{\pageref{firstpage}--\pageref{lastpage}} \pubyear{2005}

\maketitle

\label{firstpage}

\begin{abstract}

We present a comprehensive study of X-ray emission and wind properties
of {\changed massive} magnetic early B-type stars. Dedicated
\xmm\ observations were obtained for three early type B-type stars,
\xicma, \voph, and \zcas\ {\changed with recently discovered magnetic
  fields}. We report the first detection of X-ray emission from
\voph\ and \zcas.  The latter is one the softest X-ray sources among
early type stars, while the former is one of the X-ray faintest. The
observations show that the X-ray spectra of our program stars are
quite soft with the bulk of X-ray emitting material having a
temperature of about 1\,MK. We compile the complete sample of early
B-type stars with detected magnetic fields to date and existing X-ray
measurements, in order to study whether the X-ray emission can be used
as a general proxy for stellar magnetism. We find that the X-ray
properties of early massive B-type magnetic stars are diverse, and
that hard and strong X-ray emission does not necessarily correlate
with the presence of a magnetic field {\changed corroborating similar 
conclusions reached earlier for the classical chemically peculiar magnetic 
Bp-Ap stars.

We analyze the UV spectra of five non-supergiant B stars with magnetic
fields (\tsco, \bcep, \xicma, \voph, and \zcas) by means of non-LTE
iron-blanketed model atmospheres. The latter are calculated with the
Potsdam Wolf-Rayet (PoWR) code, which treats the photosphere as well
as the the wind, and also accounts for X-rays. With the exception of
\tsco, this is the first analysis of these stars by means of stellar
wind models.} Our models accurately fit the stellar photospheric
spectra in the optical and the UV. The parameters of X-ray emission,
temperature and flux are included in the model in accordance with
observations.  We confirm the earlier findings that the filling
factors of X-ray emitting material are very high.

Our analysis reveals that the magnetic early type B stars studied
here have weak winds with velocities not significantly exceeding
$v_{\rm esc}$. The mass-loss rates inferred from the
analysis of UV lines are significantly lower than predicted by
hydrodynamically consistent models. We find that, although the X-rays
strongly affect the ionisation structure of the wind, this effect is
not sufficient in reducing the total radiative acceleration. When the
X-rays are accounted for at the intensity and temperatures observed,
there is still sufficient radiative acceleration to drive stronger
mass-loss than we empirically infer from the UV spectral lines.

\end{abstract}


\section{Introduction}

Magnetic fields in massive stars ($M_*\msim 8 M_\odot$) have the
potential to influence stellar formation and evolution
\citep[e.g.][]{fer2009}. Although discoveries of massive star
magnetism are increasing (Hubrig \etal\
2011; Petit \etal\ 2011; Wade \etal\ 2011), our knowledge of their
existence, origin and the role they play remains limited. The routine
presence of magnetic fields in these stars is observationally neither
established nor disproved, due to the lack hitherto
of suitable diagnostic tools and observations of
adequate quality \citep[for a recent review see][]{dl2009}.
The Zeeman effect, commonly used to measure magnetic fields in
solar-type stars, is less useful in hot stars: the line broadening
by macroscopic motions exceeds the Zeeman splitting, unless the
magnetic field is extremely strong.

While direct measurements of magnetic field have been possible only
for the closest and brightest stars, there has been indirect evidence
for magnetic fields on massive stars.  The wide range of observational
phenomena, such as UV wind-line periodic variability, cyclic
variability in H$\alpha$ and He\,{\sc ii}\,$\lambda$4686, and excess
emission in UV-wind lines centered about the rest wavelength
\citep[e.g.][]{hen2005,sch2008} are commonly explained by the
influence that magnetic fields exert on stellar winds. For example,
cyclical wind line variability is likely due to wind flow that is
guided by a large scale, dipole-like magnetic field corotating with
the star. In these cases the timescale of the variability is similar
to the rotation period. Other indicators of magnetic fields may
include chemical peculiarity, certain pulsation behavior, non-thermal
radio emission, and anomalous X-ray emission.

In this paper we address the X-ray emission and wind properties of
magnetic B-type stars. Our study includes stars earlier than B2 that
have both confirmed magnetic fields and X-ray data.  {\changed Thus, our
sample includes only massive stars, with initial masses exceeding
$\sim 8\,M_\odot$. We consider the B-type stars, where magnetic
fields have been detected only recently as well as the earliest types of
``classical'' chemically peculiar Bp type stars. The latter have high
incidence of strong magnetic fields \citep[e.g.][]{bych2003}. Much
work has already been done to study the  X-ray emission from chemically
peculiar magnetic Bp and Ap stars. Using the results from X-ray surveys,
correlations between the X-ray properties of these stars and their
magnetism were investigated by \citet{drake1994} and \citet{leo1994}.
Those works considered a broader range of spectral classes, not only
B0-B2 stars as this work. The conclusion was reached that, despite some
notable exceptions, there is little evidence that the X-ray emission
from the Bp stars is different from other B-stars with similar spectral
types.  \citet{cz2007} carried out {\em Chandra} high-angular resolution
observations of a sample of late B-type and A-type stars with measured
magnetic fields in the range from 0.2-17 kG. They showed that the
existence of a magnetic field of kG strength on a late B-type or A-type
star is not necessarily a prerequisite for finding X-ray emission among
these stars. 

The main focus of this work is on $\beta$\,Cephei-type variables and a
slowly pulsating B-type star, where magnetic fields have been detected
only recently.} We performed dedicated observations with \xmm\ for
three magnetic B-stars: \xicma, \voph, and \zcas.  Two of them,
\voph\ and \zcas\ are detected in X-rays for the first time. We also
searched the X-ray archives to collect X-ray data for other magnetic
early B-type stars. Thus, at the time of writing, we have assembled a
complete sample of early-type magnetic B-stars for which X-ray
observations are available. This sample will, undoubtedly, grow in
future.

To infer stellar and wind parameters for the five stars in our sample we
employ the non-LTE stellar atmosphere model PoWR
\citep[e.g.][]{hg2003}.

Our sample of B-stars stars with directly measured magnetic fields is
biased to non-supergiant stars. Our program stars allow us to explore
the parameter space comprising different wind momenta, field strengths,
and rotation periods with regard to the observed X-ray emission.  We
compare the properties of X-ray emission  between our program stars and
other B-type stars of similar spectral types, but without confirmed
magnetic field. This allows investigation of whether the X-ray emission
can serve as an indicator of the surface stellar magnetic field.

In the absence of a magnetic field, the mechanism that produces
X-ray emission in early-type stars is thought to be the line-driven
instability (LDI) that is an intrinsic property of stellar winds
\citep{lucy1970}.  The LDI mechanism generates numerous
shocks in a stellar wind, where plasma can be heated up to X-ray emitting
temperatures \citep[e.g.][]{feldmeiera1997,feldmeierb1997}.

In addition to the LDI mechanism, X-rays are thought to arise when
magnetic fields are capable of governing the wind streams.  Knowledge
of the wind parameters, magnetic field topology and strength,
together with X-ray observations provides a good basis to review
critically the models of X-ray generation in massive magnetic stars.

\citet{gh1982} proposed a scenario to explain observational phenomena
associated with the B2Vp $\sigma$\,Ori\,E. Their model predicts the
formation of a torus of matter around the star, formed by a weak
stellar wind channeled along magnetic field lines.  The torus is
magnetically coupled to the star. Its co-rotation explains the
observed photometric and spectroscopic variability.  The X-ray
emission is produced when the wind streams from opposing magnetic
hemispheres collide leading to strong shock heating of the plasma.
\citet{tow2007} further developed the \citet{gh1982} scenario by using
semi-analytical models of the rigidly rotating magnetosphere (RRM) in
the limit of very strong ($\sim$few~\,kG) magnetic fields. Their model
predict very hard X-ray emission with temperatures up to 100\,MK. 

\citet{bma1997} envisage a star with a dipole magnetic field that is
sufficiently strong to confine the stellar wind. Collision between the
wind components from the two hemispheres in the closed magnetosphere
leads to a strong shock. The X-ray emission from the Ap-type star
IQ\,Aur was explained in the framework of this ``magnetically confined
wind shock model'' (MCWS), and the presence of a magnetic field in the
O-type star $\theta^1$\,Ori\,C was postulated \citep{bmb1997}. The
direct confirmation of the magnetic field in this star by
\citet{don2002} proved that X-rays have diagnostic potential in
selecting massive stars with surface magnetic fields.  Numerical MHD
simulations in the framework of the MCWS model were performed by
\citet{ud2002}. These simulations compare well with the X-ray
observations of $\theta^1$\,Ori\,C \citep{gag2005}; however, the model
has difficulties in describing the X-ray emission from other magnetic
O-type stars \citep{naze2010}. In this paper we discuss the
applicability of the MCWS model to the early-type magnetic B~stars.

\citet{cas2002}, \citet{mah2003}, \citet{jcb2008}, and \citet{mc2009}
studied the case of fast rotating magnetic massive stars with an
increasing degree of model sophistication to address the formation of
disks in classical Be-type stars. They showed that magnetic torquing
and channeling of wind flow from intermediate latitudes on a B~star
could, for plausible field strengths, create a dense disk a few
stellar radii in extent in which the velocity is azimuthal and of
order the local Keplerian speed. Unlike all the other
magnetically-guided wind scenarios mentioned, the disk model proposed
by \citet{jcb2008} is in a steady state, wind inflow being offset by a
very slow outflow across reconnecting field lines. \citet{li2008}
proposed a model, where the X-rays are produced by wind material that
enters the shocks above and below the disk region.  The model by
\citet{li2008} predicts a relation between the X-ray luminosity
normalized to the stellar bolometric luminosity (\Lx/\Lbol) and the
magnetic field strength in Be-type stars.  Compared to the Be-type
stars, the stars in our sample are slower rotators and have lower wind
density.

\begin{table*}
\caption{Magnetic early B-type stars with available X-ray observations}
\label{tab:bstar}
\medskip
\begin{tabular}{lrccccccc} \hline\hline
Name & HD & Sp   & $B^{\rm a}$  & $v\sin{i}$ & $P_{\rm rot}$  &  Dipole &
Obliquity$^{\rm b}$  & Ref \\
    &     &  & G & km\,s$^{-1}$& d &  & $\beta$ & \\
\hline
\tsco\ & 149438 &B0V  & $\langle \sim 500 \rangle$ &   5  & 41.033 & no
  &  & {\it 1}\\
\hline
\multicolumn{9}{c}{$\beta$\,Cep-type and SPB-type stars } \\ \hline
\xicma & 46328 & B0.7IV & $5300\pm 1100$ & $9 \pm 2$ & 2.18
& yes & 79.1 & {\it 2} \\
\bcep\  & 205021 &B2III     & $360\pm 40$ & $27\pm 2$ &12.00089& yes &
 $85\degr\pm 10\degr$ & {\it 3, 4} \\
V2052\,Oph & 163472 & B1V & $250\pm 190$ & $60\pm 4$ & 3.63883 & yes &
 $35\degr\pm 18\degr$  & {\it 5} \\
$\zeta$\,Cas  &3360  & B2IV & $335^{+120}_{-65}$ & $17\pm 3$ &
5.37045 & yes & $77\degr\pm 6\degr$ & {\it 6} \\ \hline
\multicolumn{9}{c}{Peculiar B stars} \\ \hline
NU\,Ori & 37061 & B0V(n)  & $\sim$620 &  $225 \pm 50$ &  & yes
&    & {\it 7} \\
V1046\,Ori & 37017 & B2V & $\langle \sim 1500 \rangle$ & $\lsim$95 & 0.9 & ? &
$42\degr$--$59\degr$ &  {\it 8, 9, 13} \\
HR\,3089 & 64740  & B1.5Vp & $\langle 572\pm 114 \rangle$ & 160 & 
1.33 & ? & & {\it 9, 13, 14}\\
LP\, Ori & 36982 & B2Vp &   $\sim$1100 & $80\pm 20$ &  & yes &  &
{\it 7} \\
$\sigma$\,Ori\,E & 37479 & B2Vp &   $\sim$10000 & $140\pm 10$ & 1.191 &
yes & $66\degr $ & {\it 10} \\
HR\,5907 & 142184  & B2.5Ve & $\sim$20000 & 280 & 0.5083 &
  yes(?) & $4\degr$ (?) & {\it 11, 12} \\
\hline  \hline
\vspace{0.1cm}
\\
\multicolumn{9}{l}{\changed
$^{\rm a}$ for dipole field configuration, B is the polar
field strength. }\\
\multicolumn{9}{l}{For \tsco, HR\,3089, and V1046\,Ori, 
an approximate average field strength is shown in angle brackets} \\
\multicolumn{9}{l}{
$^{\rm b}$ $\beta$ is the angle between the magnetic and rotational
axes for the dipole field configuration} \\
\multicolumn{9}{l}{{\it References:}
{\it 1} \citet{dtsco2006, sota2011};
{\it 2} \citet{hub2011};
{\it 3} \citet{tel1997};
{\it 4} \citet{don2001};}\\
\multicolumn{9}{l}{
{\it 5} \citet{neinervoph2003}
{\it 6}  \citet{nzcas2003};
{\it 7} \citet{pet2008};
{\it 8} \citet{boh1987};}\\
\multicolumn{9}{l}{
{\it 9} \citet{rk2008};
{\it 10} \citet{rei2000}; 
{\it 11} \citet{abt2002};
{\it 12} \citet{grun2010};}\\
\multicolumn{9}{l}{
{\it 13} \citet{bych2003};
{\it 14} \citet{bor1979}}\\
\end{tabular}
\end{table*}

The paper is organized as follows.  Our program stars are introduced
in Sect.\,\ref{sec:stars}. The new \xmm\ observations are described in
Sect.\,\ref{sec:obs}, and analyses of X-ray spectra presented in
Sect.\,\ref{sec:xspec}.  The X-ray properties of early-type Bp stars
are discussed in Sect.\,\ref{sec:pec}. Analysis of UV spectra for our
program stars is presented in Sect.\,\ref{sec:wind}.  We discuss our
results in Section\,\ref{sec:disc}, with a summary of the main points
given in Section\,\ref{sec:con}.

\section{The program stars}
\label{sec:stars}

Our {\changed sample of early type magnetic B-stars with recently
  discovered magnetic fields} includes the B0 type star \tsco, the
$\beta$\,Cephei type variables \bcep, \xicma, and \voph, and the
slow-pulsating B-type star (SPB) \zcas\ (see Table\,\ref{tab:bstar}).

\tsco\ is a well studied object, which has been observed in X-rays
by all major X-ray missions. A complex magnetic field topology was
discovered in \tsco\ by \citet{dtsco2006}. \tsco\ is a source of
hard X-ray emission and displays narrow emission line profiles,
suggesting nearly stationary plasma \citep{mewe2003}.  Recently,
\tsco\ was monitored throughout its rotational cycle by the {\em
Suzaku} X-ray observatory.  Contrary to expectations, rotational
modulations of X-ray emission were not detected \citep{ign2010}.
Magnetic fields were detected on two other stars, HD\,66665
and HD\,63425, that are spectroscopically similar to \tsco\
\citep{pet2011}.  Although it remains to compare their X-ray
characteristics to \tsco, the new discoveries suggest that \tsco\
may be a prototype for a wider class of stars.

\bcep, $\xi^1$\,CMa and V2052\,Oph are $\beta$\,Cephei type variables.
$\beta$\,Cep-type stars have masses between 7 and 16\,\Msun\ and are on
the main-sequence or in an early post-MS evolutionary phase (corresponding to
spectral type O9 to B3). Stars of this type display photometric and
spectral variations caused by pulsations in low-order pressure and
gravity modes of short-period (3--8\,h)  \citep{dz1993}.

A magnetic field in \bcep\ was detected by \citet{hen2000}. Based on
the analysis of the {\em Rosat} measurements, \citet{don2001}
suggested that the X-rays from \bcep\ can be explained within the
framework of the MCWS model. This model predicts rotationally
modulated, strong and hard X-ray emission originating at a few stellar
radii from the surface close to the disk around the magnetic
equator. However, data appear not to support these predictions. X-ray
multiphase high-spectral resolution observations of \bcep\ by
\xmm\ and {\em Chandra} fail to show either rotational modulations or
modulations on the time-scale of the stellar pulsations
\citep{fav2009}. The X-rays do not originate further out in the wind
than 2-3\,$R_\ast$, and the emission is quite soft with the bulk of
X-ray emitting plasma at T$\approx$3\,MK, \citep{fav2009}.  The X-ray
luminosity of $\beta$\,Cep is not atypical of stars with the same
spectral type.

\citet{hub2006} and \citet{sil2009} measured magnetic fields in a
sample of eight $\beta$\,Cep-type stars.  Both these studies agree in
the lack of a definite detection of magnetic fields in the six
\bcep-type stars, with typical longitudal field formal errors of few
tens of G.  The results of these investigations seems to indicate that
the presence of a strong magnetic field is not a general intrinsic property of
this spectral class.

Magnetic field measurements of \xicma\ were reported in \citet{hub2006}
and confirmed by \citet{sil2009}. Recently, \citep{hub2011} established
the magnetic field configuration, and the rotational periods for a
number of magnetic B-stars, among them \xicma. It was shown that
we see \xicma\ nearly rotational pole-on.

\xicma\ was previously observed in X-rays by {\em Einstein} and
{\em Rosat}. \citet{cas1994} presented a {\em Rosat} PSPC spectrum
of this star. They reported a temperature of 3.7\,MK and an emission
measure EM$\sim 10^{54}$ cm$^{-3}$, and pointed out that \xicma\ has the
highest X-ray luminosity and the hardest spectrum among
the \bcep-type stars in their sample.

The detection of a magnetic field in \voph\ was reported by
\citet{neinervoph2003}.  The polar magnetic field was determined
assuming that the star is an oblique rotator. \citet{neinervoph2003}
note that \voph\ is very similar to \bcep. They suggested that based
on \bcep, the wind of \voph\ should be confined and form a disk-like
structure.  Prior to our observations, there was no positive detection
of X-ray emission from \voph, with the {\em Rosat All Sky Survey}
(RASS) yielding only an upper limit to its X-ray luminosity.

The slow-pulsating B~stars are slightly less luminous and cooler
than the $\beta$\,Cep-type stars.  These stars show multi-periodic
brightness and color variations on a time-scale of 0.8\,d--3\,d.
\citet{nzcas2003} reported the discovery of a magnetic field on the
SPB star \zcas.  \citet{hub2006} and \citet{sil2009} searched for
magnetic fields in a large sample of SPB stars; however, the results
of their measurements agree only partially -- both studies detect
a magnetic field on 16\,Peg (B3V) (\citet{sil2009} report a marginal
detection).  16\,Peg is not yet detected in X-rays.  Prior to
our observations, there were no positive detection of X-ray emission
from \zcas\ either, with the RASS yielding an upper limit to its
X-ray luminosity.  \citet{nzcas2003} analyzed the magnetic field
configuration of \zcas\ and concluded that the star is an oblique
magnetic dipole.  However, they argued that there is no evidence
for a disk around this star, based on the lack of an IR excess.

\section{Observations}
\label{sec:obs}

We obtained dedicated \xmm\ observations of \xicma, \zcas, and \voph.
All three (MOS1, MOS2, and PN) European Photon Imaging Cameras (EPICs)
were operated in the standard, full-frame mode and a thick UV filter
\citep{mos2001,pn2001}. The log of observations is shown in
Table\,\ref{tab:obs}. The data were analyzed using the software {\sc
sas}\,9.0.0. Each of the stars in our sample was detected by the
standard source detection software.  The exposure times and EPIC PN
count rates for our program stars are given in Table\,\ref{tab:obs}. The
spectra and light-curves were extracted using standard procedures. The
spectra of all sources were extracted from regions with diameter
$\approx 15''$. The background areas were chosen in nearby areas free
of X-ray sources. The EPIC PN spectra of \xicma\ were corrected for
``out-of-time events''.


\begin{table}
\begin{center}
\caption{\xmm\ observations of three magnetic early B-type stars}
\vspace{1em}
\renewcommand{\arraystretch}{1.2}
\begin{tabular}[h]{lccl}  \hline
\hline Star & MJD   &  useful exposure & PN count-rate$^{\rm a}$ \\
            &       &  [ksec]         & [s$^{-1}$] \\
\hline
\xicma\ & 55046.4576 & 6.9  & $0.61 \pm\ 0.01$     \\
\voph\  & 55077.0906 & 9.2  & $0.003 \pm\ 0.001$ \\
\zcas\  & 55046.4576 & 12.3 & $0.040 \pm\ 0.002 $  \\
\hline
\multicolumn{4}{l}{$^{\rm a}$ in the 0.3-7.0\,keV band; background subtracted}\\

\end{tabular}
\label{tab:obs}
\end{center}
\end{table}
%

\section{X-ray spectra and their fitting}
\label{sec:xspec}

To analyze the spectra we used the standard spectral fitting software
{\sc xspec} \citep{arnaud1996}. The {\changed reference abundances
  were set to solar values according to \citet{aspl2009}}. The number
of counts per bin in the spectra of \zcas\ and \voph\ is small,
therefore we used Cash-statistic \citep{cash1979} to fit their
spectra, while the \xicma\ spectra were fitted using the
$\chi^2$-statistics. Using the neutral hydrogen column density as a
fitting parameter does not yield a sensible constraint on its value.
Therefore, $N_{\rm H}$ was fixed at its interstellar value as found
from the analysis of optical and UV spectra (see
Section\,\ref{sec:wind}) for all our program stars.

The standard model {\em vapec}, which assumes that the plasma is in
collisional ionization equilibrium (CIE), was used to model the
observed X-ray spectra of our program stars.  CIE models require that
several equilibria exist in the stellar wind namely that the electrons
and ions are both thermalized and have equilibrated so as to have
$T_{\rm e} = T_{\rm i}$ and that ionization and recombination rates
are balanced.  The thermalization processes is controlled by Coulomb
collisions or even faster plasma effects.  However, cooling of the
plasma can be faster than the overall recombination time due to of
some ions with important cooling lines. \citet{sd1993} provide a
comparison of the collisional equilibrium timescale and the cooling
timescale for carbon and iron ions. From their Fig.\,15, at
temperatures above $\msim$1\,MK, the cooling time is $\tau_{\rm
  cool}=1.5nkT/n_{\rm e}n_{\rm i}\Lambda(T) > 5\times 10^8/n_{\rm
  e})$\,hr, and the collisional equilibrium time for C\,{\sc vi} is
$\tau_{\rm recomb}=\alpha^{-1}_{\rm recomb}\cdot n_{\rm e}^{-1} < 1.5\times
10^8(1\,{\rm cm}^{-3}/n_{\rm e})$\,hr, where $n_{\rm e}$ and $n_{\rm i}$
are electron and ion number densities respectively in cm$^{-3}$,
$\Lambda(T)$ is the cooling function, and $\alpha_{\rm recomb}$ is the
recombination coefficient (see \citet{sd1993} for details).  Hence, at
the temperatures where the maximum of emission measure is determined
from the analysis of X-ray spectra (see Table\,\ref{tab:xspecmod}),
the cooling of the plasma is slower than the recombination time-scale,
and the conditions for the establishing the CIE seem to be fulfilled.

Using \xicma\ as an example, we deduce from our wind models (see
Section\,\ref{sec:wind}) that at 1\,$R_\ast$ distance from the
photosphere, the electron density in the wind is $\sim
10^7$\,cm$^{-3}$. The cooling time of a 1\,MK plasma is $\approx
50$\,hr, while the collisional equilibrium time for C\,{\sc vi} is
$\approx 15$\,hr. Note that this is long compared to the pulsational
time-scale of $\beta$\,Cep-type variables, e.g.\ the pulsational
period of \xicma\ is 5.03\,hr \citep{st2005}.

The mean free path for a Coulomb collision $l_{\rm mfp}\approx 9.4
\times 10^{7} T^2 n_8^{-1}$\,cm, where $T$ is in the units of MK, and
$n_8$ is in the units of $10^8$\,cm$^{-3}$. Using parameters for
\xicma\, close to the stellar surface, the value of $l_{\rm mfp}$ is only
100\,cm. Therefore, we believe that the use of CIE models to describe
the X-ray spectra of our program stars is justified.

The chemical abundances of our program stars deviate from the solar,
which is most likely a consequence of the surface magnetic field.
Peculiar abundances are often found in magnetic stars, and is
commonly explained by diffusion processes which allow heavier
elements to sink in the atmosphere under the influence of gravity,
while lighter elements are lifted to the surface by radiation
pressure.  The summary of the relevant publications with analyses
of abundances in our program stars can be found in \citet{neinervoph2003},
\citet{nzcas2003} and \citet{mor2008}.  In summary, these studies
agree that the magnetic stars show an overabundance of nitrogen,
and sometimes helium. The quality of the X-ray spectra are not
sufficient to constrain abundances.  Therefore, for each star we
have used the abundances determined by \citet{mor2008}.

\begin{figure}
\centering
\includegraphics[height=0.99\columnwidth, angle=-90]{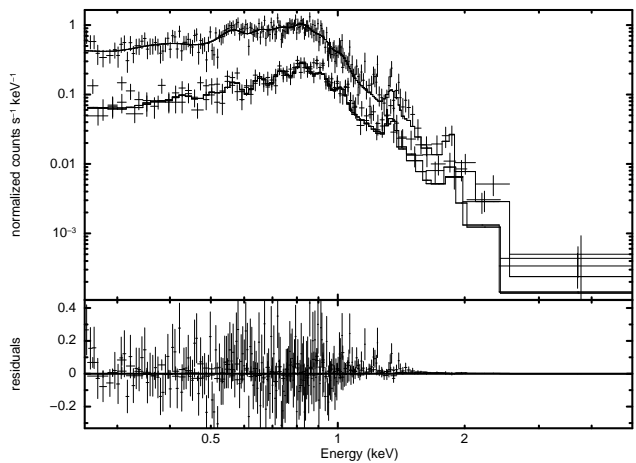}
\caption{ \xmm\ PN (upper curve), and MOS1 and MOS2 (lower curves)
spectra of \xicma\ with the best fit
three-temperature model (solid lines). The model parameters are
shown in Table\,\ref{tab:xspecmod}.}
\label{fig:xisp}
\end{figure}

\medskip
\noindent
\underline{\xicma} The observed EPIC spectra of \xicma\ and the fitted
model are shown in Fig.\,\ref{fig:xisp}. The abundances were set to
solar values, except C/C$_\odot$=0.56, N/N$_\odot$=1.48, and
O/O$_\odot$=0.79, as found by \citet{mor2008}. We initially attempted
to fit the spectra using two temperature models similar to
\citet{fav2009} for modeling the \bcep\ X-ray spectrum.  However, we
found that adding a softer model component with $kT\approx 0.1$\,keV
significantly improved the quality of the fit. Parameters of the
best-fit model are shown in Table\,\ref{tab:xspecmod}. We also
attempted to fit the \xicma\ spectrum assuming a higher temperature
component.  By analogy with the \tsco\ spectral models
\citep{mewe2003}, a spectral model component with $kT_4=1.7$\,keV was
assumed. However, the emission measure of this hot component cannot be
constrained. As a next step, we investigated whether the
\xicma\ spectrum can be fitted with non-equilibrium ionization
models. Using the {\em nei} model available in {\sc xspec}, we failed
to find a suitable set of model parameters to reproduce the observed
spectrum.

%

\medskip \noindent \underline{\voph}  Our \xmm\ observation detected the
X-ray emission from \voph\ for the first time. The source has a very low
count rate (see Table\,\ref{tab:obs}); in total only about 60 source
counts were collected.  At a distance of $d=417$\,pc, the X-ray
luminosity of  \voph\ is only $\Lx\approx 3\times
10^{29}$\,erg\,s$^{-1}$,  making it one of the least X-ray luminous
early type B-stars in the sky. To model the observed low S/N spectrum,
we fixed  the neutral hydrogen column density at its interstellar value.
The PN spectrum of \voph\ can be well described using a two temperature  
plasma model see  Table\,\ref{tab:xspecmod}). The abundances C/C$_\odot$=0.6,
N/N$_\odot$=1.4, and O/O$_\odot$=0.5 were used as found by
\citet{mor2008}. The temperature and the emission measure of soft
component are only poorly constrained but the presence of a
soft component is required to reproduce the observed hardness ratio.
There are no indications of a harder component being present in the
spectrum of \voph.  \\

Pausing to consider the X-ray spectral analyses of the \bcep-type
stars (\bcep, \xicma\, and \voph), it emerges that the spectral
temperatures are similar among these stars. We define the mean
spectral temperature $\left< kT \right>$ as the emission measure
weighted average temperature, with

\begin{equation}
\left< kT \right> \equiv \sum_i
kT_i\cdot {\rm EM}_i / \sum_i {\rm EM}_i.
\end{equation}

\noindent Then, the mean spectral temperature for all magnetic
$\beta$\,Cep-type stars is the same at $\approx$3.5\,MK (see
Table\,\ref{tab:xspecmod}). However, we find a large difference in
X-ray luminosities for the \bcep\ stars, with \xicma\ being 4 times
more X-ray luminous than \bcep\ and 50 times more X-ray luminous
than \voph\ (see Table\,\ref{tab:bstar}).

\begin{figure}
\centering
\includegraphics[height=0.99\columnwidth, angle=-90]{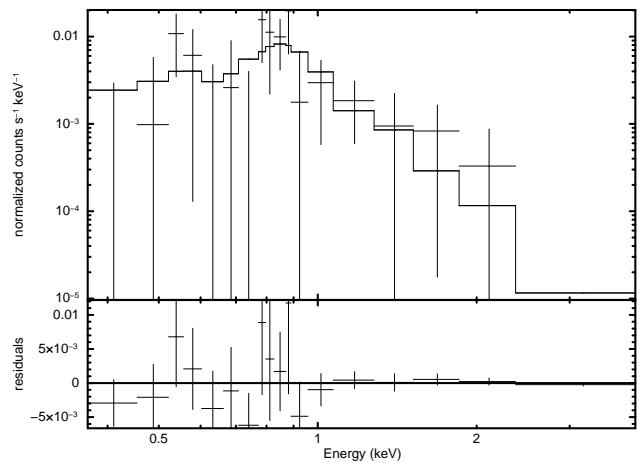}
\caption{ \xmm\ EPIC-PN spectrum of \voph\
and the best fit two-temperature model. The model parameters are
shown in Table\,\ref{tab:xspecmod}.}
\label{fig:vophsp}
\end{figure}
%

\begin{figure}
\centering
\includegraphics[height=0.99\columnwidth, angle=-90]{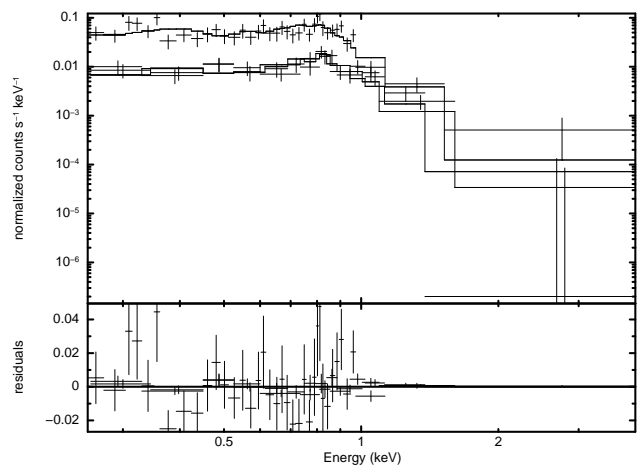}
\caption{ \xmm\ EPIC PN (upper), and MOS1 and MOS2 (lower)
spectra of \zcas\ and the best fit two-temperature
model. The model parameters are shown in Table\,\ref{tab:xspecmod}.}
\label{fig:zcassp}
\end{figure}

\medskip
\noindent
\underline{\zcas} Our \xmm\ observations of \zcas\ are the first X-ray
observations to yield a spectrum of a magnetic SPB-star.  Visual inspection 
of EPIC \xmm\ images already reveals that \zcas\ is a very
soft X-ray source. This expectation is confirmed by the X-ray
spectral fits. The EPIC spectra of \zcas\ and the best fit 2T CIE model are
shown in Fig.\,\ref{fig:zcassp}. A solar composition was assumed,
except for C/C$_\odot$=0.5, N/N$_\odot$=1.4, and O/O$_\odot$=0.5
\citep{mor2008}.  Parameters of the model are shown in
Table\,\ref{tab:xspecmod}.  The maximum temperature inferred from the
spectral analysis is $\approx$4\,MK, and there are no indications of
a harder spectral component. The emission measure is dominated by the
1\,MK plasma; the hotter, 4\,MK component constitutes less than 20\%
of the total emission measure. \zcas\ has the softest X-ray spectrum
among all hot stars where magnetic field have been detected. The mean X-ray
spectral temperature of \zcas\ is about 1\,MK (see
Table\,\ref{tab:xspecmod}).

%


\begin{table*}
\begin{center}
\caption{The spectral parameters derived from the \xmm\ EPIC
observations of our program stars assuming the multi-temperature
CIE plasma models ({\em vapec}) corrected for the interstellar absorption
({\em tbabs}). The values which have no error have been frozen during
the fitting process. The corresponding spectral fits are shown in
Figs.\,\ref{fig:xisp},\ref{fig:zcassp},\ref{fig:vophsp}. For comparison, the
spectral parameters inferred from modeling \xmm\ data of
\bcep\ and {\em Suzaku} spectra of \tsco\ are also shown.
\label{tab:xspecmod}}
\vspace{1em}
\renewcommand{\arraystretch}{1.2}
\begin{tabular}[h]{lccccc}  \hline
\hline
Star                          & \xicma\          &  \zcas\          & \voph\
& \bcep$^{\rm a}$ & \tsco$^{\rm b}$\\ \hline
$N_{\rm H}^{\rm c}$ [$10^{20}$\,cm$^{-2}$] & 1.4 & 3 & 15 & $2.5\pm 0.1$ & 3 \\ \hline
$kT_1$ [keV]                  & $0.12\pm 0.01$   & $0.08\pm 0.02$  & $0.14 \pm 0.12$
& $0.24 \pm 0.01$ & $0.11 \pm 0.01$\\
EM$_1$ [$10^{53}$cm$^{-6}$]   & $22.48 \pm 5.64$ & $1.29\pm 1.05$  & $0.006\pm 0.025$
& $11 \pm 2$      & $17.0 \pm 2.61$\\ \hline
$kT_2$ [keV]                  & $0.32\pm 0.01$   & $0.31 \pm 0.02$ &
&                 & $0.34 \pm 0.01$\\
EM$_2$ [$10^{53}$cm$^{-3}$]   & $ 19.3 \pm 3.36$ & $0.27\pm 0.07$  &
&                 & $10.4 \pm 0.51$\\ \hline
$kT_3$ [keV]                  & $0.68\pm 0.05$   &                 & $0.65 \pm 0.11$
& $0.69\pm 0.03$  & $0.71 \pm 0.10$\\
EM$_3$ [$10^{53}$\,cm$^{-6}$] & $6.41 \pm 2.57$  &                 & $0.003\pm 0.002$
& $1.3 \pm 0.3$   & $7.2\pm 0.3$   \\ \hline
$kT_4$ [keV]                  &                  &                 &
&                 & $1.52 \pm 0.06$\\
EM$_4$ [$10^{53}$\,cm$^{-6}$] &                  &                 &
&                 & $5.2\pm 0.3$   \\ \hline
$\left< kT \right>\equiv \sum_i kT_i\cdot {\rm EM}_i / \sum_i {\rm EM}_i$\,[keV]
    & 0.3 & 0.1 & 0.3 & 0.3 & 0.5 \\ \hline
Flux$^{\rm d}$ [10$^{-12}$\,erg\,cm$^{-2}$\,s$^{-1}$] & 1.1 & 0.093 & 0.006 & 1.0 & 16.4 \\
 \hline \hline
\multicolumn{5}{l}{$^{\rm a}$ the values are adopted from Favata \etal\ (2009)} \\
\multicolumn{5}{l}{$^{\rm b}$ the values are adopted from Ignace \etal\ (2010)} \\
\multicolumn{5}{l}{$^{\rm c}$ correspond to the ISM hydrogen column
density for all stars} \\
\multicolumn{5}{l}{$^{\rm d}$ in the 0.3-7.0\,keV band, except of \tsco\ in
the 0.3-10.0\,keV band,
absorbed; } \\
\end{tabular}
\end{center}
\end{table*}

\section{\changed X-ray properties of classical peculiar magnetic 
early-type B stars}
\label{sec:pec}

{\changed In this section we briefly consider the X-ray properties of
  chemically peculiar Bp-stars, addressing only stars with early
  spectral types. The classical chemically peculiar Bp-Ap stars have a
  high incidence of strong surface magnetic fields.  The observed
temporal variations of longitudal magnetic fields are usually
compatible with dipole or low-order multipole fields inclined to the
rotation axis \citep{dl2009}.  The strong magnetic fields strongly
influence weak stellar winds, making the Bp~stars excellent test cases
for models of X-ray production in magnetic early-type stars.

There has been earlier work seeking to establish the X-ray properties
of chemically peculiar stars.  \citet{drake1994} searched the RASS
database at the positions of about 100 magnetic Bp-Ap stars. They
detected 10 X-ray sources and argued that in four cases the X-ray
emission presumably arises from an early-type star with a radiatively
driven wind. The X-ray luminosities were found to be in general
agreement with the canonical values of X-ray emission from massive
stars $L_{\rm X}\approx 10^{-7}\,L_{\rm bol}$. E.g.,\ the B1.5Vp star
HR\,3089 has $\log{L_{\rm X}/L_{\rm bol}}\approx -7.2$.}

{\changed To include newer data,} we selected Bp stars with spectral
types earlier than B2 from {\it The Catalog of Peculiar Magnetic Stars}
\citep{rk2008}.  This sample was augmented by two more magnetic Bp-type
stars reported by \citet{pet2008} and the magnetic Be star HR\,5907
recently reported by the MiMeS collaboration \citep{grun2010}. As a next
step we searched the available X-ray catalogs. Our search revealed that
$\sigma$\,Ori\,E, LP\,Ori, NU\,Ori, and V1046\,Ori, HR\,5907 have
positive X-ray detections.

$\sigma$\,Ori\,E is the prototypical oblique fast magnetic rotator. Its
X-ray emission is quite hard: the EM of 10\,MK plasma is as large as the
EM of the cooler plasma with T$\approx 3$\,MK \citep{sf2004}. The mean
spectral temperature of $\sigma$\,Ori\,E (in quiescence) is $0.66$\,keV
or 7.6\,MK. {\changed We estimate the bolometric luminosity of
$\sigma$\,Ori\,E, $\log{\Lbol/L_{\odot}}=3.8$ , based on its spectral
type and UBV colours and assuming a distance of 640\,pc
\citep{hunger1989}.  The average (quiescent+flare) X-ray luminosity in
0.1-2.4\,keV band at 640\,pc is $L_{\rm X}\approx 8\times
10^{31}$\,erg\,s$^{-1}$, the quiescent X-ray luminosity is factor of 5
lower than in the peak \citep{sf2004}}. The star is X-ray luminous: its
ratio of X-ray to bolometric luminosities \Lx/\Lbol\ is nearly two
orders of magnitude higher than normally found in B-type stars (see
Table\,\ref{tab:xray}).  An extremely rare event in the history of X-ray
observations of massive hot stars -- an X-ray flare -- has been observed
from $\sigma$\,Ori\,E \citep{gs2004,sf2004} that may be understood as a
result of centrifugal breakout of the torus \citep{mul2009}.  These
X-ray properties of $\sigma$\,Ori\,E seems to confirm the expectations
of the RRM model \citep{tow2007}.

Given the general nature of these models, one would naturally expect
that these results could be applied to other BV stars with strong
magnetic fields.  This however, seems not to be the case.

The young star LP\,Ori is similar to  $\sigma$\,Ori\,E in many
respects: it has a similar age, spectral type, and a kG strong
magnetic field. It is also a  source of hard X-ray emission
\citep{get2005}. However, the X-ray luminosity of LP\,Ori is
significantly lower than the X-ray luminosity of $\sigma$\,Ori\,E.
In fact, the X-ray luminosity of LP\,Ori is nearly the lowest one
among all magnetic early B-stars (see Table\,\ref{tab:xray}). The
models predicts modulations of the X-ray luminosity due to the
occultation of X-ray emitting site by the opaque torus. These
modulations are thought to occur on a time-scale comparable to the
rotational period of the star. LP\,Ori was observed by {\em Chandra}
for $\approx$8.8\,d \citep{get2005}. This long exposure time suggests
that the observed X-ray flux is somewhat averaged over the rotational
period, and the star is indeed intrinsically faint in X-rays.

X-ray observations of $\sigma$\,Ori\,E and LP\,Ori seemingly suggest
that the strong magnetism may account for the hardness of the
X-ray emission. However, the Bp stars V1046\,Ori and NU\,Ori provide
counter examples.

Similar to $\sigma$\,Ori\,E, V1046\,Ori has a kG magnetic field. It
also is an oblique magnetic rotator, whose obliquity and rotational
period are similar to that of $\sigma$\,Ori\,E.  The star was
serendipitously observed by {\em XMM-Newton}. No X-ray spectra are
available, but hardness ratios are provided in {\it The Second
  XMM-Newton Serendipitous Source Catalog or 2XMMi-DR3} and in
\citet{naze2009}. We have compared these hardness ratios with those of
\tsco\ reported in the same catalog.  V1046\,Ori is a softer X-ray
source compared to \tsco, and its X-ray luminosity is relatively low
(see Table\,\ref{tab:xray}). {\changed We estimate its bolometric
  luminosity based on UBV colors and spectral type}.

The magnetic field on NU\,Ori is only about $\sim$0.5\,kG and is
weaker than the fields of the other Bp stars in our sample.
\citet[and references therein]{pet2008} comment on NU\,Ori being a
triple system, containing a massive B0.5V primary, along with a low
mass spectroscopic companion and a visual companion. {\changed NU\,Ori
  was recently analyzed by \citet{sim2011}, who obtained
  $\log{\Lbol/L_{\odot}}=4.4$. }  NU\,Ori was observed by {\em Chandra}
  simultaneously with LP\,Ori \citep{get2005}.  Surprisingly, that
  analysis indicates that the X-ray luminosity of NU\,Ori is a factor
  of 30 higher than for LP\,Ori, yet the former has a considerably
  softer emission as compared to the latter.

HR\,5907 has very strong magnetic field \citep{grun2010}. Assuming a
dipole, the field strength at the pole is $\approx 20$\,kG. The star is
a fast rotator with $v\sin{i}\approx 280$\,km\,s$^{-1}$ \citep{abt2002}.
\citet{yud2001} detect intrinsic polarization on the level 0.17\%\ in V
band and consider this star as a classical Be star with a disk. The star
was observed by {\em Rosat} PSPC for $\approx 4.8$\,ks. It is detected
with a flux $\approx 2\times 10^{-13}$\,\flux\ corresponding to rather
low X-ray luminosity (see Table\,\ref{tab:xray}). We fit the low
signal-to-noise {\em Rosat} PSPC spectra and find that HR\,5907 has a
relatively hard spectrum characterized by a temperature above 1\,keV.

The rigidly rotating magnetosphere scenario makes robust predictions of
strong, variable, and hard X-ray emission.  It appears that in five out
of the six early-type magnetized B stars under discussion, these
predictions are not fulfilled. Considering just our small sample of
early-type B stars with kG strong magnetic fields, {\changed we must
conclude that the new measurements confirm earlier results on the X-ray
emission from Bp-Ap stars. The properties of their X-ray emission are
quite diverse, with majority of the stars emitting X-rays on the level
typical for the general sample of B-stars \citep[e.g][]{drake1994}.}  

{\changed Placed in the context of this earlier work, our results
  corroborate the basic conclusion that while strong and hard X-ray
  emission is sufficient to suggest a star may be magnetic, it is not a
  required property of magnetic stars. }

\begin{table}
\caption{\changed X-ray luminosities of magnetic early B-type stars} 
\label{tab:xray}
\medskip
\begin{tabular}{lccc} \hline\hline
Name & d  & \Lx\   & $\log\left(L_\text{X}/L_\text{bol}\right)$ \\
     & pc &  $10^{30}$\,erg\,s$^{-1}$ &                  \\
\hline
\tsco\ & 150 & 40 & -6.4 \\
\hline
\multicolumn{4}{c}{Magnetic $\beta$\,Cep-type and SPB-type stars } \\ \hline
\xicma\     & 420  & 30  & -6.6 \\
\bcep\      & 200 & 6.4 & -7.0 \\
\voph\      & 400  & 0.3 & -8.0 \\
\zcas\      & 180 & 0.5 & -7.5 \\ \hline
\multicolumn{4}{c}{Other magnetic early-type B stars} \\ \hline
NU\,Ori & 400  & 1.0 & -8 \\
V1046\,Ori$^{\rm a}$ & 380 & 0.1 & -8.0 \\
$\sigma$\,Ori\,E$^{\rm b}$    & 640         & 80   & -5.6 \\
{\changed HR\,3089$^{\rm c}$}   & 300         & 2    & -7.2\\
LP\,Ori             & 470         & 0.02 & -8.5 \\

HR\,5907            & 120         & 0.4 & -7.4\\
\hline  \hline
\\
\multicolumn{4}{l}{
Distances are from \citet{vanleeuwen2007} except of LP\,Ori, $\sigma$\,Ori\,E,
HR\,3089;}\\ 
\multicolumn{4}{l}{$^{\rm a}$ the distance between the X-ray source;}\\
\multicolumn{4}{l}{
2XMM\,J053522.0-042938 and the position of V1046\,Ori is $3''$;} \\
\multicolumn{4}{l}{$^{\rm b}$ \Lx\ (quiescent+flare) in 0.1-2.4\,keV
band from \citet{sf2004} ;} \\
\multicolumn{4}{l}{\changed $^{\rm c}$ \Lx\ in 0.1-2.4\,keV band from
\citet{drake1994};}\\
\end{tabular}
\end{table}

\section{Stellar winds}
\label{sec:wind}

\begin{figure*}
\centering
\includegraphics[width=0.8\textwidth]{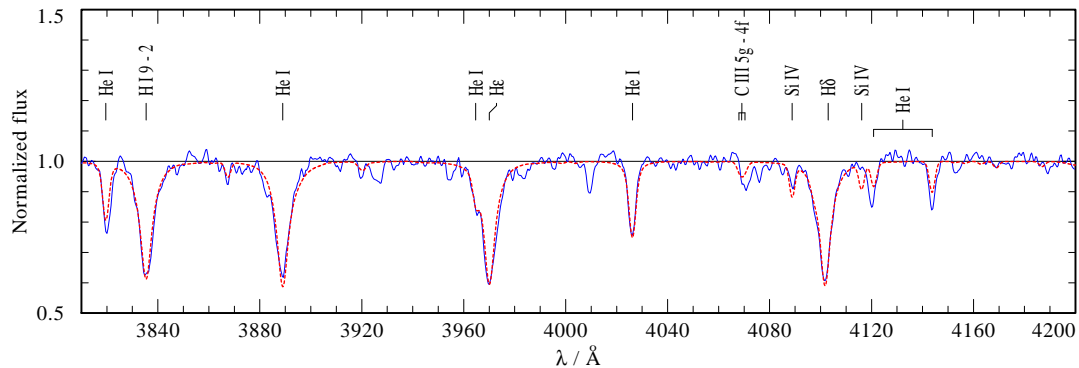}
\caption{Optical CTIO spectrum of \xicma\ obtained on 1988/11/02  
{\changed \citep{wf1990}}
(blue line) vs.\ a PoWR model spectrum (red  line) using
stellar parameters of \xicma\ as shown in Table\,\ref{tab:stellarparameters}.
The model spectrum is convolved with a 2.5\,\AA\ Gaussian to match the spectral
resolution of the ``2D-Frutti'' spectrograph at CTIO. }
\label{fig:xiop}
\end{figure*}

The strong UV radiation of B-type stars drives their stellar winds.
The parameters of stellar winds are commonly inferred from the
analyses of optical and UV spectral lines by means of stellar
atmosphere models. While the B-type giants and supergiants have often
been analyzed, the winds of BIV and BV type stars yet remained
relatively unexplored. \citet{pr1989} used profile fits to model
spectral lines of C\,{\sc iv}, Si\,{\sc iv}, and Si\,{\sc iii} in
high-resolution spectra of B~stars {\changed obtained with the
  International Ultraviolet Explorer (IUE)}, and produced a
homogeneous set of wind-velocity and column-density measurements for
40 non-emission, non-supergiant B~stars. It was concluded that the
presence of C\,{\sc iv} resonance doublet in their UV spectra must be
due to the effect of X-rays on wind ionization. This effect is often
referred to as ``superionization''. \citet{pr1989} did not find any
asymmetry in the Si\,{\sc iv} lines which would be typical if the
stellar wind was strong.  Estimates of column densities revealed that
the mass-loss rates in non-supergiant B stars are significantly lower
than in O~and Be~stars.  From fitting the UV resonance lines of
C\,{\sc iv} and Si\,{\sc iv}, it was found that the products of
mass-loss rate and ionization fraction, $\mdot q$, are
$\ll$$10^{-10}$\,\myr. Regarding the wind velocities, it was shown
that the winds of ``normal'' non-supergiant stars are slow compared to
those of the Be~stars. The maximum observed wind velocities do not
generally exceed the stellar escape velocity.

{\changed In this paper we analyze  the UV spectra of our sample
  magnetic stars by means of a stellar atmosphere code. Stellar
  spectroscopy codes used to analyze massive stars are based on
  spherically symmetric or plane-parallel geometries, despite of the
  clear observational evidence that massive star winds can depart from
  spherical symmetry \citep[e.g.][]{kap1997, ham2001}.  It is very
  probable that magnetic fields affect the wind geometry. However, at
  present 3D modeling of stellar winds that would allow quantitative
  spectroscopic analysis is beyond reach.  Our approach allows to
  conduct a  multiwavelength X-ray and UV/optical spectral
  analysis. The obtained wind parameters of our sample stars can be
  compared to other massive stars in the general framework of massive
  star wind quantitative spectroscopy.}

\begin{table}
\begin{center}
\captionabove{UV observations of program stars used in the analysis}
\begin{tabular}{lcc}
\hline \hline
object & data set & date \\
\hline
$\zeta$ Cas  & SWP\,51309  & 04/07/1994 \\
$\xi^1$ CMa  & SWP\,03574 & 12/12/1978  \\
$\tau$ Sco   & SWP\,55997 & 23/09/1995 \\
V\,2052\,Oph & SWP\,50431 &  31/03/1994 \\
$\beta$ Cep  & SWP40477  &  28/12/1990 \\
\hline
\end{tabular}
\label{tab:iue}
\end{center}
\end{table}

\subsection{Model description: PoWR stellar atmosphere}

In this work we analyze the spectra of magnetic B~stars by means of
stellar atmosphere models that account for the presence of X-rays.  We
use the Potsdam Wolf-Rayet (PoWR) model atmosphere code \citep{hg2003,
  hg2004}.  The PoWR code has been used extensively to analyze not
only massive stars with strong stellar winds \citep[e.g.][]{lier2010},
but also low-mass central stars of planetary nebulae, and extreme
helium stars \citep{hamann2010, htodt2010}. The PoWR code solves the
non-LTE radiative transfer in a spherically expanding atmosphere
simultaneously with the statistical equilibrium equations and accounts
at the same time for energy conservation.  Complex model atoms with
hundreds of levels and thousands of transitions are taken into
account. The computations for the present paper include complex atomic
models for hydrogen, helium, carbon, oxygen, nitrogen, and silicon.
Iron and iron-group elements with millions of lines are included in
the PoWR code through the concept of superlevels \citep{graf2002}.
The extensive inclusion of the iron group elements is important not
only because of their blanketing effect on the atmospheric structure,
but also because the diagnostic wind lines in the UV (e.g.\ the
C\,{\sc iv} and Si\,{\sc iv} resonance lines) are heavily blended with
the ``iron forest''.

The PoWR models can take account of stellar-wind clumping in the
standard volume-filling factor `microclumping' approach
\citep[e.g.][]{hk1998}, or even, with an approximate correction for
wind clumps of arbitrary optical depth
\citep[``macroclumping'',][]{osk2007}. However, since the wind
signatures are quite weak for all stars of our sample, and since there
is no way to estimate the degree of clumping, we apply here only
smooth-wind models. Note that microclumping does not directly
influence the resonance line profile shapes via radiative transfer
effects, but can indirectly affect the lines by virtue of modifying
the ionization stratification.

The X-ray emission and its effects on the ionization structure of the
wind is included in the PoWR atmosphere models according to the recipe
of \citet{baum1992}. We assume an optically thin, hot plasma component
distributed within the ``cool'' stellar wind.  The uniform value of
the so-called Emission Measure 'filling factor' , $X_{\rm
  fill}=EM_{\rm hot}/EM_{\rm cool}$ (really a density squared weighted
volume ratio), is adjusted such that the emergent X-ray luminosity
agrees with the observations.  The X-ray emissivity is restricted to
the fast-wind domain, for which we assume a minimum radius of
1.1\,\Rstar.  The absorption of the X-ray radiation by the ``cool''
stellar wind is taken into account, as well as its effect on the
ionization stratification by the Auger process.

The lower boundary of the model atmosphere is set at a Rosseland
depth of 20, meaning that the (nearly) static photosphere is included
in the computations.  The velocity field consists of two parts. In
the photospheric part of the atmosphere, a hydrostatic density
stratification is assumed, while for the wind the usual ``$\beta$-law''
prescription is adopted, in our case with an exponent $\beta=1$.
We also tested the so-called double-$\beta$ law \citep{hilliermiller1999},
but without substantial improvement of the fit quality.

Each stellar atmosphere model is defined by the effective temperature,
surface gravity, luminosity, mass-loss rate, wind terminal velocity
\vinf, and chemical composition. The gravity determines the density
structure of the stellar atmosphere below and close to the sonic
point. From the pressure-broadened profiles of photospheric lines,
the spectroscopic analysis allows derivation of the gravity and thus
the stellar mass. For our sample of B stars, we confirm a 
discrepancy between the  spectroscopic  and evolutionary mass, the
former being lower than the latter (see \citet{wv2010} and
references therein).

Dedicated analyses of the photospheric spectra of our program stars
have been performed by \citet{neinervoph2003}, \citet{nzcas2003} and
\citet{mor2008}. Using their abundances, $T_{\rm eff}$, and $\log g$
we find that our model reproduces photospheric spectra very well, as
demonstrated by the example shown in Fig.\,\ref{fig:xiop}. Therefore, to
speed up the analysis, we adopt the literature values of $T_{\rm eff}$,
and $\log g$. The effects of rotation with values $v\sin{\rm i}$ from
Table\,\ref{tab:bstar} are also included in the models.

\begin{figure*}
\centering
\includegraphics[width=0.9\textwidth]{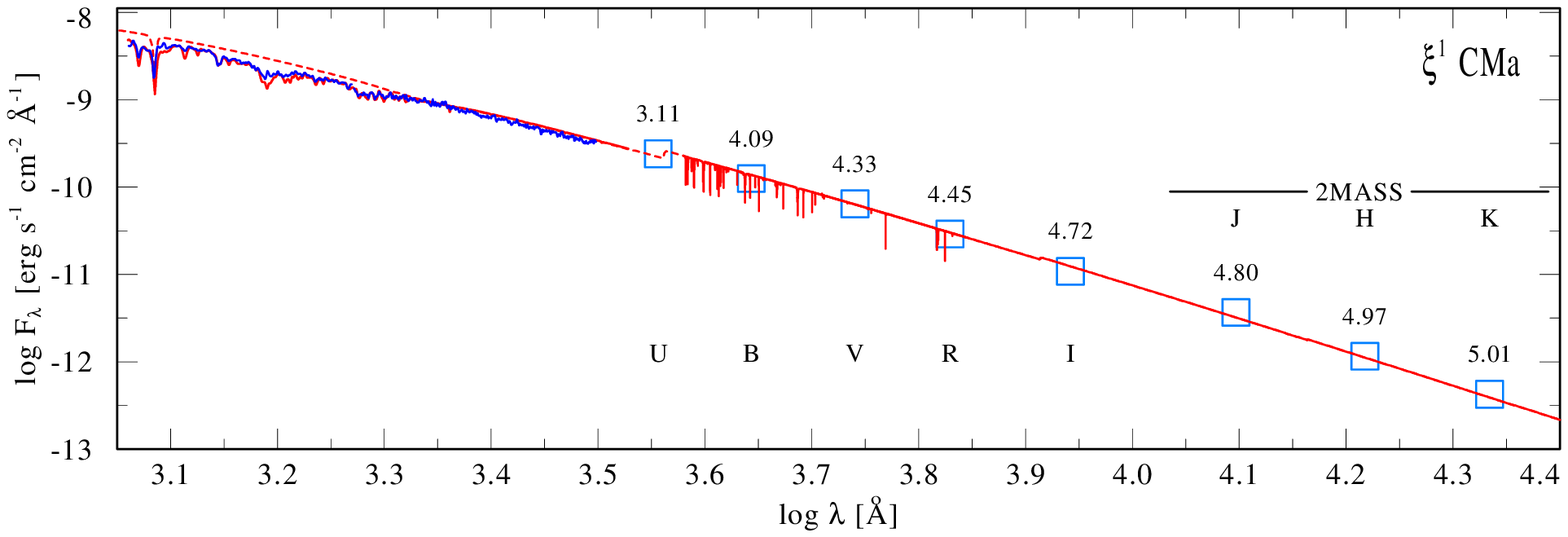}
\caption{Spectral energy distribution for \xicma. Observed IUE spectra
are shown as thin blue lines. Blue boxes indicate observed photometry
(labels: magnitudes) taken from the 2MASS catalogue
\citep{skrutskie2006} and from \citet{morelMagnenat1978}. The synthetic
spectrum (red line) is calculated using parameters given in
Table\,\ref{tab:stellarparameters}. The model flux is reddened with
$E_{{\rm B}-{\rm V}}=0.04$ and  corrected for interstellar Lyman line
absorption. The model continuum without lines (red dotted) is also
shown to demonstrate how the iron group lines form a pseudo-continuum in
the UV.}
\label{fig:xised}
\end{figure*}
%

All stars in our sample have measured parallaxes. Hence we can
scale synthetic spectral energy distributions (SED) to that distance
and fit to the  observations, covering the whole wavelength range
from the UV to IR.  Furthermore, the model spectra are corrected
for interstellar extinction. Dust extinction is taken into account
using the reddening law of \citet{cardelli1989}.   An  example of
a SED fit is shown in Fig.\,\ref{fig:xised}. The figure also shows
the model continuum without lines, illustrating how the ``forest''
of iron lines forms a pseudo-continuum in the UV that is well
reproduced by our model.

With $\log g$, $T_{\rm eff}$, and $\Lbol$ being fixed, we compare
the synthetic and observed lines varying the wind parameters $\mdot$
and $v_\infty$ in order to achieve the best fit.  The usual indicators
of mass loss used in B~supergiants and Be~stars, as well as in
O-type stars, are H$\alpha$ together with the UV resonance lines.
However, in our sample of non-supergiant stars the H$\alpha$ line
is entirely photospheric\footnote{\bcep\ shows H$\alpha$ emission
episodes.  However, \citet{schnerr2006} found that the H$\alpha$
emission is not related to the primary in $\beta$\,Cep, but is due
to its 3.4 mag fainter companion that is a classical Be star.}.
Therefore, our wind diagnostic can only be based on the UV resonance
lines. High resolution IUE spectra were retrieved from the archive
for all stars of our sample. For some of them, multiple IUE
observations are available \citep[see e.g.\ the IUE time series for
\bcep\ and \xicma\ in][]{sch2008}. In those cases we selected
the observation of best quality (cf.\ Table\,\ref{tab:iue}).


\subsection{$\tau$ Sco}


\begin{figure}
\centering
\includegraphics[width=0.9\columnwidth]{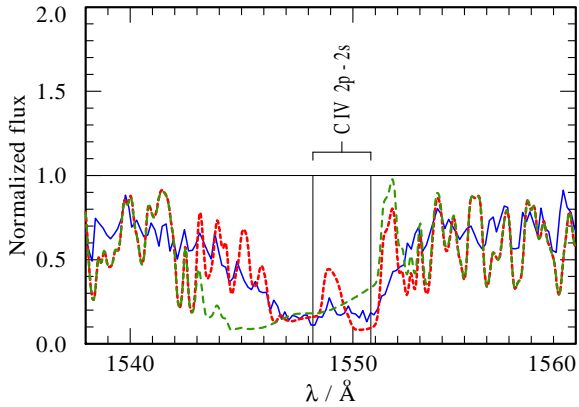}
\caption{ $\tau$ Sco: The effect of ionization by X-rays on C\,{\sc iv}
 $\lambda\lambda$1548.2,\,1550.8\,\AA\ doublet.
Detail of the UV spectrum observed with IUE (blue
  thin line) vs.\ PoWR models: without X-rays (green dotted line) and
  with X-rays (red thick line). The model parameters:
  $\log(\dot{M})=-9.3$, $v_\infty = 1000\,\text{km/s}$. This figure
  shows that C\,{\sc iv} is efficiently destroyed by X-rays in
  the outer parts of the atmosphere. Without accounting for the
  ionization by X-rays, the mass-loss rate would be underestimated.
  \label{fig:tau_sco-CIV-withoutXrays}}

\end{figure}

\begin{figure}
\centering
\includegraphics[width=0.9\columnwidth]{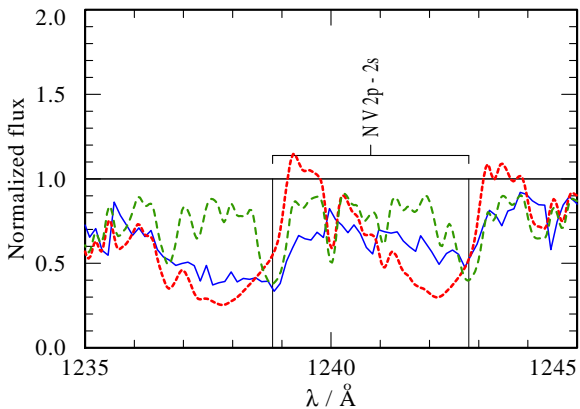}
\caption{$\tau$ Sco: Same as in Fig.\,\ref{fig:tau_sco-CIV-withoutXrays}
but centered on the N\,{\sc v}  doublet.}
\label{fig:tau_sco-NV-withoutXrays}
\end{figure}

\begin{figure}
\centering
\includegraphics[width=0.9\columnwidth]{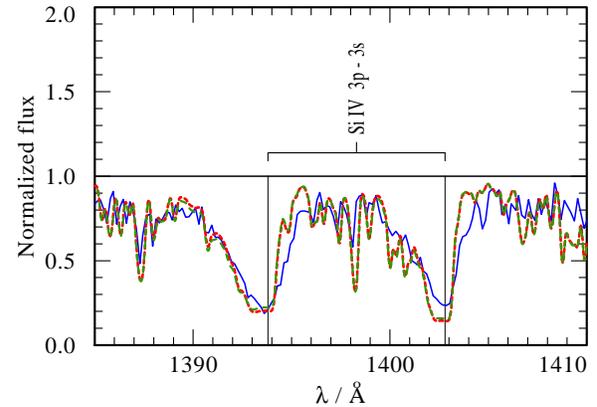}
\caption{$\tau$ Sco: Same as in Figs.\,\ref{fig:tau_sco-CIV-withoutXrays}
and \ref{fig:tau_sco-NV-withoutXrays} but centered on the Si\,{\sc iv} doublet.
The model parameters: $\log(\dot{M})=-8.6$, $v_\infty = 1000\,\text{km/s}$.
Note that even without X-rays,
  most silicon is
  already ionized to Si\,{\sc v} in the outer parts of
the atmosphere, therefore the Si\,{\sc iv} doublet
  is insensitive to X-rays.
  \label{fig:tau_sco-SiIV-withoutXrays}
 }
\end{figure}

\tsco\ is a well studied object that has remained one of the primary targets
in stellar UV and X-ray astronomy from their early days
\citep[e.g.][]{rog1977,mac1989}.  It was one of the first stars whose
UV spectrum was analyzed by means of atmosphere models based on
co-moving radiative transfer techniques by \citet{wrh1981}, who
obtained a mass-loss rate $\log\mdot=-8.9\pm 0.5$ from the analysis
of UV lines. It was also shown that O\,{\sc vi} and N\,{\sc v} lines
cannot be reproduced by the cool wind models.
\citet{co1979} explained the presence of O\,{\sc vi} and N\,{\sc v} as
arising from Auger ionization by X-rays.

\begin{figure}
\centering
\includegraphics[width=0.9\columnwidth]{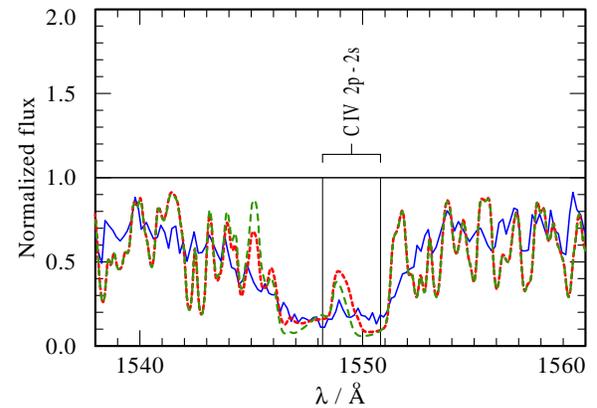}
\caption{$\tau$ Sco: Determination of the wind velocity from modeling
the C\,{\sc iv} line. Detail of the UV spectrum observed with IUE (blue
  thin line) vs.\ PoWR models with $v_\infty=500$\,km\,s$^{-1}$ (green
  dotted line) and $v_\infty=1000$\,km\,s$^{-1}$ (red dotted line). The
  derived mass-loss rate is $\log(\dot{M})=-9.3$. The blue absorption wing of the
  C\,{\sc iv} line is better matched by the model with higher
  $v_\infty$. Both models include ionization due to X-rays with
  parameters as listed in Table \ref{tab:bstar}.
       \label{fig:tau_sco-CIV-twovelos}
 }
\end{figure}

We re-analyzed the wind properties of \tsco\ using the PoWR model.
The abundances as presented in \citet{hub2008} were adopted.
We obtained wind parameters by modeling the C\,{\sc iv}, N\,{\sc v},
and Si\,{\sc iv} lines. During our analysis we found that the line
profiles in the IUE range are strongly influenced by a combination
of three parameters at the same time: the terminal wind velocity
\vinf, the mass-loss rate \mdot, and the impact of superionization
via X-ray emission.  It is not possible to disentangle these effects
and therefore our solution may not be unique.

\begin{figure}
\centering
\includegraphics[width=0.9\columnwidth]{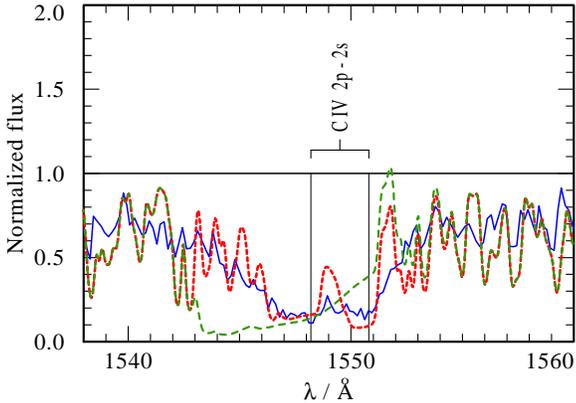}
\caption{$\tau$ Sco: Constraining the mass-loss rate from modeling of
the C\,{\sc iv} doublet.  Detail of the UV spectrum observed with IUE (blue
  thin line) vs.\ PoWR models with $\log{\mdot}= -8.6$ (green
  dotted line) and  $\log{\mdot}= -9.3$ (red dotted line).
Both models use $v_\infty =
        1000\,\text{km/s}$ and include superionization.
        The
        C\,{\sc iv} doublet is better matched by the model with the
        lower $\dot{M}$.
        \label{fig:tau_sco-CIV}}
\end{figure}

We include the X-ray flux and temperature as deduced from the
observations.  The need to account for X-rays in wind modeling can
be nicely demonstrated using C\,{\sc iv} and N\,{\sc v} doublets
(Figs.\,\ref{fig:tau_sco-CIV-withoutXrays} and
\ref{fig:tau_sco-NV-withoutXrays}). Although it is possible to
reproduce the observed C\,{\sc iv} doublet without X-ray superionization
by assuming lower $\dot{M}$ and lower $v_\infty$, this method fails
for the N\,{\sc v} doublet. As the effective temperature of B~stars
is not high enough to create sufficient amounts of N\,{\sc v} by
photoionization, the observed P~Cygni line profile of the N\,{\sc
v} doublet can only be reproduced if superionization due to X-rays
is included. On the other hand, the Si\,{\sc iv} doublet in \tsco\
has little sensitivity to the X-rays
(Fig.\,\ref{fig:tau_sco-SiIV-withoutXrays}).

With fixed parameters for the X-ray emitting gas from this study,
the terminal velocity of the wind can be deduced from
fitting the line profiles. Fig.\,\ref{fig:tau_sco-CIV-twovelos}
shows the asymmetric line profile of the C\,{\sc iv}-line doublet
at $\lambda\lambda\,1548.2,\,1550.8$ \AA.  We calculated three
models taking $v_\infty=500$,\, km\,s$^{-1}$, $1000$\,km\,s$^{-1}$
and $1500$km\,s$^{-1}$ and keeping other parameters the same. Although a 
higher wind velocity cannot be excluded, we find that a model
with $v_\infty=1000$\,km\,s$^{-1}$ described the observed lines
best.


\begin{figure}
\centering
\includegraphics[width=0.9\columnwidth]{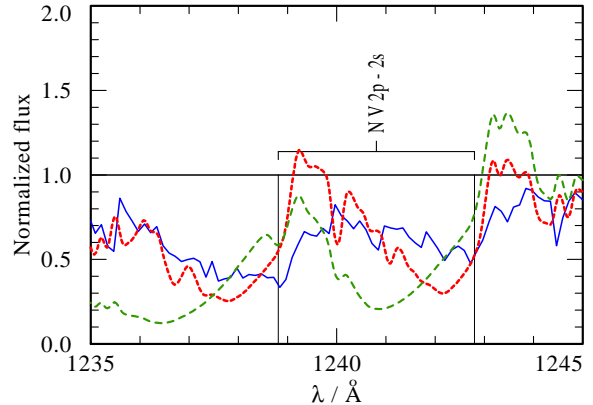}
\caption{ $\tau$ Sco: Same as in Fig.\,\ref{fig:tau_sco-CIV} but
for the N\,{\sc v} doublet. The P\,Cygni line profile of
  the N\,{\sc v} doublet is strongly affected  by X-ray ionization.
The observed line is better fitted by a model with a low mass-loss 
rate  $\log{\mdot}= -9.3$ (red dotted line).
\label{fig:tau_sco-NV}
}
\end{figure}

\begin{figure}
\centering
\includegraphics[width=0.9\columnwidth]{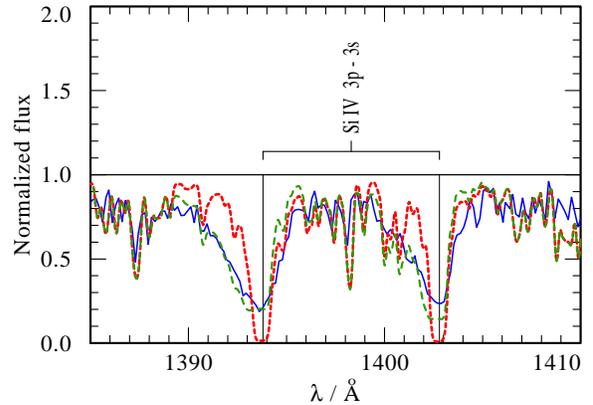}
\caption{$\tau$ Sco:
Same as in Fig.\,\ref{fig:tau_sco-CIV} and  Fig.\,\ref{fig:tau_sco-NV}  but
for the Si\,{\sc iv} doublet. In contrast to the C\,{\sc iv} and N\,{\sc v}
doublets,  the observed doublet of Si\,{\sc iv}  is better matched
by a model with a higher mass loss of $\log{\mdot}=-8.6$.}
\label{fig:tau_sco-SiIV}
\end{figure}

Setting $v_\infty = 1000\,$km\,s$^{-1}$, attention was next directed
toward a determination of the mass-loss rate. Models with mass-loss
rates in the range $\log(\dot{M} / M_\odot {\rm
yr}^{-1})=\mbox{-9.3}\,\ldots\,\mbox{-8.6}$ were computed and
compared to the UV data. We were not able to find a unique solution
which could describe all lines equally well. The C\,{\sc iv} doublet
is better reproduced with lower $\mdot=10^{-9.3}$\,\myr\ as illustrated
in Fig.\,\ref{fig:tau_sco-CIV}.  This mass-loss rate is also favored
by the model fitting of the N\,{\sc v} doublet (see
Fig.\,\ref{fig:tau_sco-NV}).  On the other hand, Si\,{\sc iv} doublet
is best described by a model with a higher mass loss of
$\mdot=10^{-8.6}$\,\myr\ as shown in Fig.\,\ref{fig:tau_sco-SiIV}.

The steps as described above were performed for all our program stars.

\subsection{\bcep}

\begin{figure}
\centering
\includegraphics[width=\columnwidth]{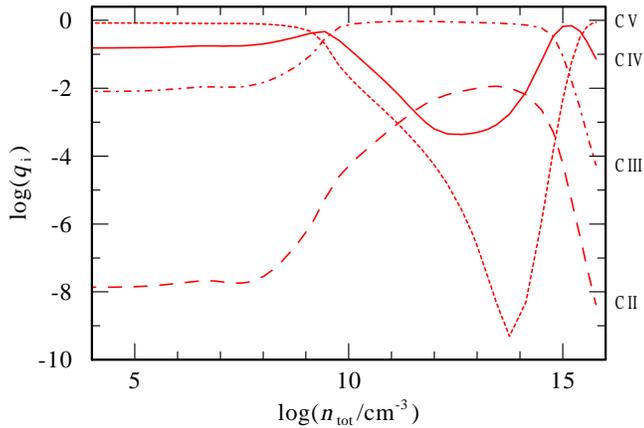}
\caption{Relative ionization fractions of carbon in the wind of \bcep\
as function of density in the wind, with declining density
corresponding to larger radius.
The wind model is calculated with $v_\infty=700$\,km\,s$^{-1}$,
$\log{\mdot}=-9.4$; stellar parameters are from
Table\,\ref{tab:stellarparameters}.}
\label{fig:qC}
\end{figure}

\bcep\ is a well-studied object. Despite this, to our knowledge,
there has been no prior detailed modeling of the UV resonance lines
from this star.  \citet{don2001} adopted a terminal wind velocity
of $\vinf=900$\,km\,s$^{-1}$. Using a constraint on the product $\mdot
q$({\rm C}\,{\sc iv}) obtained by \citet{pr1989}, and adopting an
ionization fraction of C\,{\sc iv} at 0.1\%, Donati \etal\ estimated
a mass-loss rate of $\log{\mdot} = -9$ for \bcep.  They also noticed
that this mass-loss rate is significantly lower than the predicted
value of $2.4\times 10^{-8}$\,\myr\ for a CAK model of a star like
\bcep\ \citep{abb1982}.

The ionization fractions of carbon obtained by detailed modeling
of the \bcep\ atmosphere with the PoWR code are shown in
Fig.\,\ref{fig:qC}.  The effect of ionization by X-rays prevents
C\,{\sc iv} from being the dominant ion anywhere in the stellar
wind except right at the photospheric level.  Therefore systematically
lower mass-loss rates would be obtained from modeling of the C\,{\sc
iv} doublet by models that do not account for the X-rays. (The
effect is illustrated in Fig.\,\ref{fig:tau_sco-CIV-withoutXrays}
using \tsco\ as an example.)

\begin{figure}
\centering
\includegraphics[width=0.9\columnwidth]{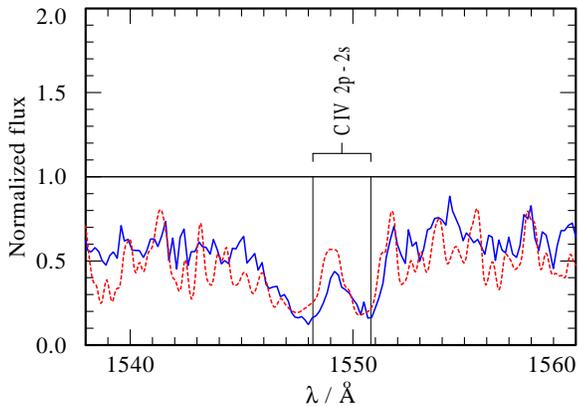}
\caption{$\beta$ Cep:  Detail of the UV spectrum observed with IUE (blue
  thin line) vs.\ PoWR models with $v_\infty =700$\,km\,s$^{-1}$ and
$\log{\mdot}=-9.4$ (red dotted line). The
  C\,{\sc iv} $\lambda\lambda$1548.2,\,1550.8\,\AA\ doublet is shown.}
\label{fig:bcep-CIV}
\end{figure}

\citet{sch2008} studied wind-line variability in magnetic B-stars.
They considered all 81 UV spectra of \bcep\ available in the IUE
archive and showed that the C\,{\sc iv} doublet is strongly modulated
with the rotation period of 12\,d.  Their Fig.\,2 shows the gradual
transition from an enhanced to a reduced contribution of emission
centered close to zero velocity in C\,{\sc iv}, as typical in
magnetic B stars.

\begin{figure}
\centering
\includegraphics[width=0.9\columnwidth]{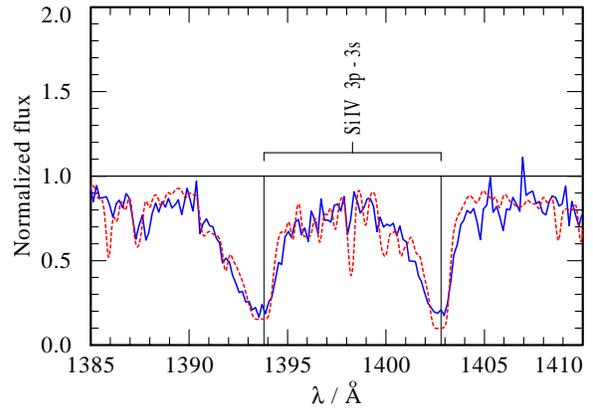}
\caption{$\beta$ Cep:
The same as in Figs.\,\ref{fig:bcep-CIV} and but for the Si\,{\sc iv}
$\lambda\lambda$1393.8,\,1402.8\,\AA\ doublet. A PoWR model with
$\log{\mdot}=-9.1$ (red dotted line) is shown. }
\label{fig:bcep-SiIV}
\end{figure}

Figure\,\ref{fig:bcep-CIV} shows the observed C\,{\sc iv} doublet
compared to a model that assumes $\vinf = 700$\,km\,s$^{-1}$ and
$\log{\mdot} = -9.4$. A higher mass-loss rate of $\log{\mdot} = -9.1$ is
required to reproduce Si\,{\sc iv} line, as shown in
Fig.\,\ref{fig:bcep-SiIV}.

\subsection{$\xi^1$ CMa}

This is the first analysis of the UV spectral lines of the
$\beta$\,Cep-type star \xicma\ by means of stellar atmosphere models.

\begin{figure}
\centering
\includegraphics[width=0.9\columnwidth]{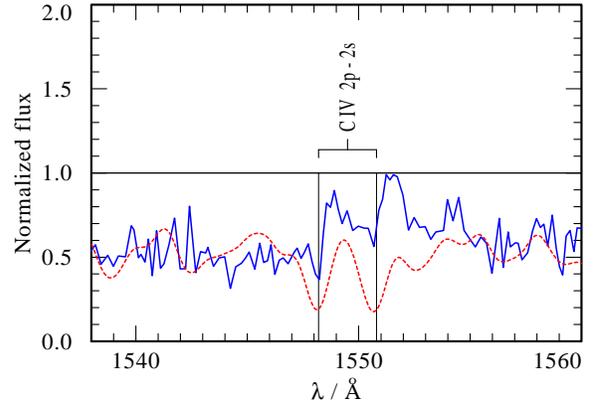}
\caption{$\xi^1$ CMa:
Detail of the UV spectrum observed with IUE (blue
  thin line) vs.\ PoWR models with $v_{\infty}=700$\,km\,s$^{-1}$ and
  $\log{\mdot}=-10$ (red dotted line). The C\,{\sc iv}
  $\lambda\lambda$1393.8,\,1402.8\,\AA\ doublet is shown. }
\label{fig:xi1_cma-CIV}
\end{figure}

In modeling the spectrum of this object, we note that the IUE data
of this star are of rather low quality.  It seems that the background
subtraction was not performed correctly, because the minimum of the
interstellar Lyman alpha absorption line is not ``black'' as it
should be. Therefore we adjusted the background level accordingly.
Our model reproduces the broad SED from the UV to the IR very well
(see Fig.\,\ref{fig:xised}). Figure\,\ref{fig:xiop} shows our model
compared to the optical spectrum of \xicma\ -- the model produces quite a 
good match to the line spectra. Therefore, it is surprising
that we were not able to achieve good quality fits to the C\,{\sc
iv} (see Figs.\,\ref{fig:xi1_cma-CIV}) and N\,{\sc v} lines.

\begin{figure}
\centering
\includegraphics[width=0.9\columnwidth]{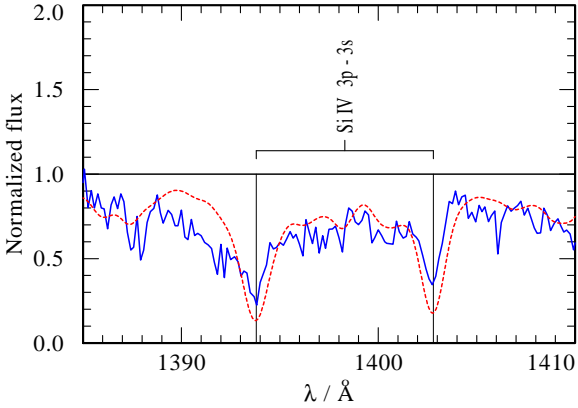}
\caption{$\xi^1$\,CMa:
The same as in Figs.\,\ref{fig:xi1_cma-CIV}
but for the Si\,{\sc iv}
  $\lambda\lambda$1393.8,\,1402.8\,\AA\ doublet.}
 \label{fig:xi1_cma-SiIV}
\end{figure}

It is tempting to suggest that the reason for these modeling
difficulties may be due to the pole-on orientation of
\xicma\ \citep{hub2011}, although the poor quality of the data is a
concern as well.  The observed line profiles of Si\,{\sc iv} (see
Fig.\,\ref{fig:xi1_cma-SiIV}) are roughly reproduced by our model.  We
adopted a larger turbulence in the wind compared to the photosphere
and convolved the UV model spectra with a Gaussian of 1\,\AA\ FWHM,
corresponding to a turbulence velocity of about 200\,km\,s$^{-1}$.

For \bcep\ and \voph, the mass-loss rate required to reproduce the
Si\,{\sc iv} doublet is higher as compared to what is needed to fit
C\,{\sc iv}, an effect that we attribute to the superionization by
X-rays. This, however, is not the case for \xicma.
Figure\,\ref{fig:xi1_cma-CIV} shows the C\,{\sc iv} line compared
to model line with $\log{\mdot}=-10$, as found from the analysis
of Si\,{\sc iv}.

\citet{sch2008} presented a time-series analysis of IUE observations
of the C\,{\sc iv} doublet in \xicma\ and noticed the lack of temporal
modulations in the spectra.  Increasing
the mass-loss rate would result in a stronger absorption feature, which
is not observed.  Therefore, we conclude that the UV lines in \xicma\
are peculiar compared to other magnetic stars in our sample, either
due to the quality of the observations or reflecting intrinsic peculiarity
to the source itself.

\subsection{\voph}
\label{sec:voph}

\begin{figure}
\centering
\includegraphics[width=0.9\columnwidth]{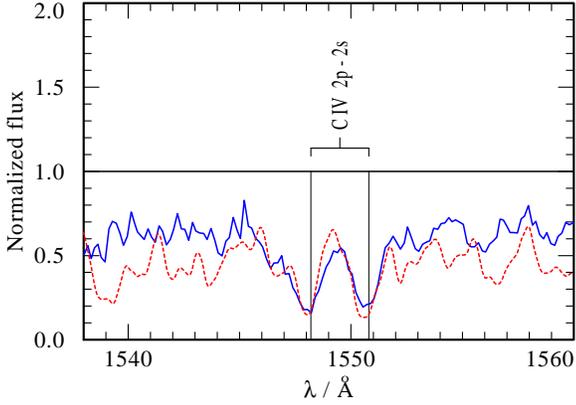}
\caption{\voph: Detail of the UV spectrum observed with IUE (blue
  thin line) vs.\ PoWR models with $v_\infty = 700$\,km\,s$^{-1}$ and
$\log{\mdot}=-10.7$ (red dotted line). The
  C\,{\sc iv} $\lambda\lambda$1548.2,\,1550.8\,\AA\ doublet is shown.}
\label{fig:v2052_oph-CIV}
\end{figure}

This is the first analysis of wind properties of \voph.  As with the 
previous stars (except \tsco), we adopt a wind terminal velocity
of $\vinf =700$\,km\,s$^{-1}$. We did calculate models with $\vinf
=1000$\,km\,s$^{-1}$, but results were similar.  Fig.\,\ref{fig:v2052_oph-CIV} 
shows the C\,{\sc iv}
doublet in the IUE spectrum of \voph\ compared to a wind model with
$\log{\mdot}=-10.4$. The effective temperature of \voph\ is relatively
low at $\Teff=23$\,kK, and C\,{\sc iii} would have been the leading
ionization stage if there were no X-ray emission. As with other
stars in our sample, the effect of superionization proves critical,
in this case elevating the ionization fraction of C\,{\sc iv} in
the wind.

\begin{figure}
\centering
\includegraphics[width=0.9\columnwidth]{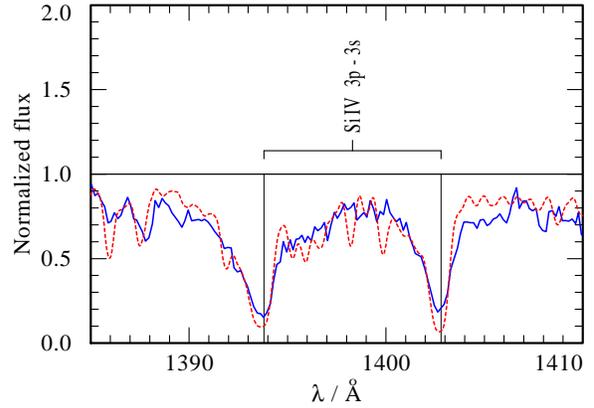}
\caption{\voph:
The same as in Fig.\,\ref{fig:v2052_oph-CIV}  but for the Si\,{\sc iv}
$\lambda\lambda$1393.8,\,1402.8\,\AA\ doublet. A PoWR model with
$v_{\infty} =700$\,km\,s$^{-1}$ and $\log{\mdot}=-9.7$ (red dotted line)
is shown. }
\label{fig:v2052_oph-SiIV}
\end{figure}

We were not able to achieve a satisfactory fit to the N\,{\sc v}
doublet with a model that includes X-ray emission. This line is
very sensitive to the presence of ionizing photons in the wind, and
even models with a mass loss as low as $\log{\mdot}=-10.7$ produces
a N\,{\sc v} doublet that is stronger than observed. This is an
interesting challenge because the observed level of X-ray emission
in \voph\ is quite low, smaller than in the other stars of our
sample.

Similarly to results for other program stars, the Si\,{\sc iv} doublet
is better reproduced with models that assume higher mass-loss rates
then required for the C\,{\sc iv} and N\,{\sc v} doublets.
Figure\,\ref{fig:v2052_oph-SiIV} shows a model fit with
$\log{\dot{M}}=-9.7$ to the Si\,{\sc iv} lines. Note that the low
$T_{\rm eff}$ of about $23$\,kK would normally suggest that the
dominant ion stage of silicon would be Si\,{\sc iv}; however, with the
presence of X-rays, Si\,{\sc v} becomes the dominant ion, and the
Si\,{\sc iv} doublet shows a more photospheric absorption profile.

\subsection{\zcas}

Ours is the first analysis of the wind properties of \zcas. This
star is the coolest among our sample. Again, we fixed the parameters
of X-ray emission based on \xmm\ data, and calculated a range of
models for various values of \vinf\ and \mdot.


\begin{figure}
\centering
\includegraphics[width=0.9\columnwidth]{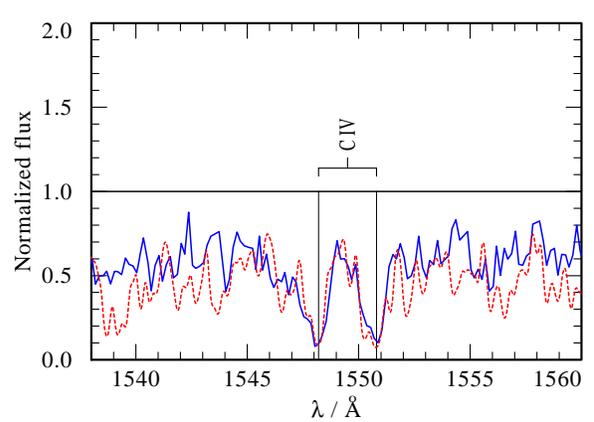}
\caption{$\zeta$ Cas: Detail of the UV spectrum observed with IUE (blue
  thin line) vs.\ PoWR models with $v_\infty =700$\,km\,s$^{-1}$ and
$\log{\mdot}=-11$ (red dotted line). The
  C\,{\sc iv} $\lambda\lambda$1548.2,\,1550.8\,\AA\ doublet shows
a purely photospheric absorption profile without any wind signature.}
\label{fig:zet_cas-CIV}
\end{figure}

Figure \ref{fig:zet_cas-CIV} shows the observed C\,{\sc iv} doublet
plotted together with a PoWR model. The observed line shows a purely
photospheric absorption profile without any wind signature and is
quite well fitted with a PoWR model. From such a photospheric
absorption profile, it is not possible to infer the wind velocity.
We have therefore assumed a terminal wind speed of $\vinf=700$\,km\,s$^{-1}$.

\begin{figure}
\centering
\includegraphics[width=0.9\columnwidth]{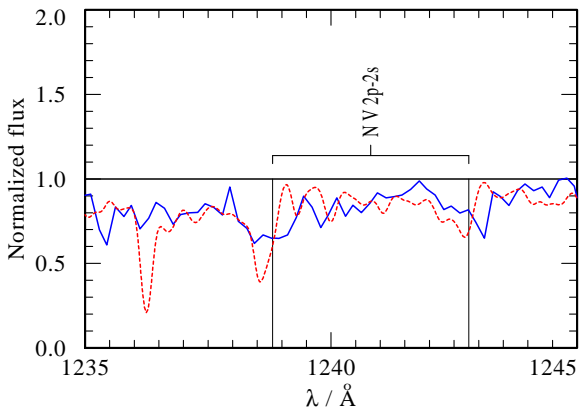}
\caption{$\zeta$ Cas: Same as in Fig.\,\ref{fig:zet_cas-CIV} but
for the N\,{\sc v}  $\lambda\lambda$1238.81,\,1242.8\,\AA\ doublet.}
\label{fig:zet_cas-NV}
\end{figure}

Wind emission signatures are present in the N\,{\sc v} doublet as
seen in Fig.\,\ref{fig:zet_cas-NV}. Note that the model N\,{\sc v}
line has a P~Cygni line profile because N\,{\sc v} becomes the
leading ionization stage via the effect of superionization. However,
in order to reproduce  this line we have to adopt a small mass-loss
rate of only $\log{\mdot} = -11$.  Surprisingly, even with this
mass-loss rate, the model line is somewhat stronger than observed.

\begin{figure}
\centering
\includegraphics[width=0.9\columnwidth]{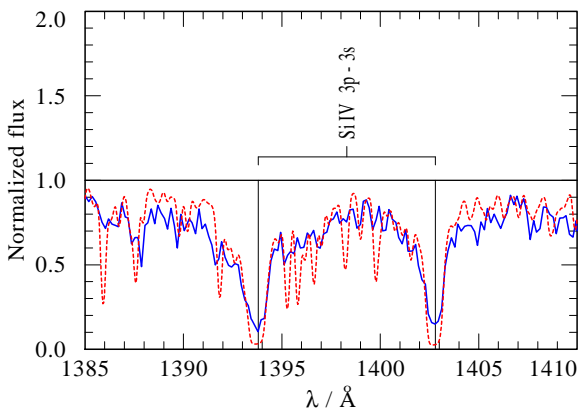}
\caption{$\zeta$ Cas:
The same as in Figs.\,\ref{fig:zet_cas-CIV} and
\ref{fig:zet_cas-NV} but for the Si\,{\sc iv}
$\lambda\lambda$1393.8,\,1402.8\,\AA\ doublet. A PoWR model with
$v_{\infty} =700$\,km\,s$^{-1}$ and $\log{\mdot}=-9.7$ (red dotted line)
is shown. }
\label{fig:zet_cas-SiIV}
\end{figure}

Similarly to other stars, there is no unique solution for
the mass-loss rate, with different UV lines indicating somewhat
different mass-loss rates. Figure\,\ref{fig:zet_cas-SiIV} shows
models of Si\,{\sc iv} compared with the IUE data.  At a low $T_*$
of about $21$\,kK, Si\,{\sc iv} would be the dominant ion stage,
but X-ray emissions make Si\,{\sc v} the dominant ion in the wind
so that the Si\,{\sc iv} line shows a more photospheric absorption
profile.

\subsection{Uncertainties on \mdot\ determination in non-supergiant B
stars}

\begin{table*}
\begin{center}
\captionabove{
Stellar and wind  parameters of \tsco, \bcep, \voph, \xicma, and \zcas.
The highest and the lowest mass-loss rates \mdot\ as obtained from the
modeling of C\,{\sc iv} and Si\,{\sc iv} doublets for each
star are given. The \mdot\ value for \xicma\  is as obtained from the
model of the Si\,{\sc iv} doublet only (see Section\,5.4 for details).
The  work ratio $Q$ (see Eq.\,\ref{eq:q} for definition) is computed for
the highest mass-loss rate that  is given in this table.}
\begin{tabular}{lccccccccccc}
\hline \hline
Star & $E_{{\rm B}-{\rm V}}$ & $T_\text{eff}$ & $\log L$ & $\log g$  &
$R$ & \multicolumn{2}{c}{$\log\dot{M}$}  &

$v_\infty$ & $v_\text{esc}$
 & Work Ratio $Q$  & $\log\left(\frac{L_\text{X}}{L_\text{bol}}\right)$ \\
    &   & kK & $L_\odot$ & cm\,s$^{-2}$  & $R_\odot$ &
     \multicolumn{2}{c}{[\myr]}  & km\,s$^{-1}$& km\,s$^{-1}$ & & \\ \hline
  &   &      &  &  & & 
  C\,{\sc iv} & Si\,{\sc iv}    &  & \\
\hline

\tsco\  & 0.03   & 30.7 & 4.3   & 3.97  & 5.0 & -9.3 & -8.6 & 
1000 & 810 & 2 & $-6.4$ \\

\bcep\  & 0.03   & 25.1 & 4.2 & 3.62 & 7.0  & -9.4 &  -9.1 &  
700  & 640 & 3 & $-7.0$  \\

\xicma\ & 0.04    & 27.0 & 4.5   & 3.7  & 8.2 & & -10 &  
700  & 750 & 14 & $-6.6$ \\

\voph\  & 0.26  & 23.0 & 3.9 & 4.0   & 5.7 & -10.7 & -9.7 &  
700  & 900 & 3 & $-8.0$ \\

\zcas\  & 0.04   & 20.9 & 3.7   & 3.7  & 5.4 & -11.0 & -9.7 & 
700  & 620 & 2 & $-7.5$  \\

\hline\hline

\end{tabular}
\label{tab:stellarparameters}
\end{center}
\end{table*}


The results of the analysis of UV and optical spectra of magnetic
B stars are summarized in the Table\,\ref{tab:stellarparameters}.
There is a discrepancy between modeling the C\,{\sc iv} and N\,{\sc v}
doublets versus the Si\,{\sc iv} doublet, therefore we determine the
mass-loss rates in our program stars only to within a factor of a
few. 

The C\,{\sc iv}, N\,{\sc v}, and Si\,{\sc iv} doublets differ in their
sensitivity to stellar atmosphere parameters. Our models show that
C\,{\sc iv} can be the leading ionization stage in the hottest stars
in our sample, such as \tsco\ and \xicma. Then X-rays destroy C\,{\sc
  iv} by photoionization (see
Fig.\,\ref{fig:tau_sco-CIV-withoutXrays}). \citep[In the context of
  dwarf O-stars this was discussed by][]{mar2005, marcolino2009}.  In
cooler stars with $T_{\rm eff} < 27$\,kK, the leading ionization stage
of carbon in the wind becomes C\,{\sc iii}. In this case Auger
ionization by X-rays leads to the creation of C\,{\sc v}, followed by
recombination to C\,{\sc iv}. Therefore, the C\,{\sc iv} line is quite
sensitive to both the stellar radiation field as determined by $T_{\rm
  eff}$ and to the X-ray emission.

N\,{\sc v} is not produced in the winds of stars with $T_\ast<
50$\,kK \citep[see Fig.\,3 in][]{hg2004}. All stars in our sample
are significantly cooler, therefore the N\,{\sc v} line in their
winds originates exclusively from the ionization by X-rays.  For
stellar temperatures below 27\,kK, the leading ion in the wind is
N\,{\sc iii}, which is Auger ionized by X-rays to N\,{\sc v}. At
27\,kK and above, the leading ion is N\,{\sc iv}. When X-rays are
present in the wind, the leading ion becomes N\,{\sc v}, or for
\tsco\ ($T_\ast = 31$\,kK) even N\,{\sc vi}.

All models that include X-ray emission and have appreciable mass-loss
rates show N\,{\sc v} with a strong P~Cygni line profile, which is not
observed (see Figs.\,\ref{fig:tau_sco-NV} and \ref{fig:zet_cas-NV}).To
achieve agreement with the observations low mass-loss rates must be
adopted.  For the $\beta$\,Cep-type variables among out program stars,
the models reproduce the observed N\,{\sc v} doublet best if the X-ray
emission is ``switched off''. Since X-ray emission is observed, one
potential explanation for curiosity could be in assuming some form of
shielding of the cool stellar wind from the X-ray radiation.

The ionization of Si is also sensitive to stellar and X-ray radiative
field. Models predict that in the stars with $T_\ast < 27$\,kK, the
leading ion would be Si\,{\sc iv}, but the presence of X-rays make
Si\,{\sc v} the dominant ion. Interestingly in hotter stars with
photospheric temperatures above 27\,kK, the dominant ion is Si\,{\sc
v} with or without X-rays; the X-rays are not sufficient to produce
a significant amount of Si\,{\sc vi}. Let's consider \bcep\ and
\tsco\ as examples. Ignoring X-ray emission in the models, Si\,{\sc
iv} is the dominant ion in the outer wind zone, and therefore the
Si\,{\sc iv} doublet becomes asymmetric in shape.  With X-rays most
of the Si\,{\sc iv} in the outer wind is destroyed (ionized to
Si\,{\sc v}), and the remaining Si\,{\sc iv} absorption is only
photospheric. On the other hand in \tsco\ that is hotter, the
dominant ionization stage is Si\,{\sc v} and the Si\,{\sc iv} line
is not sensitive to the X-ray emission.

These considerations show that the stellar radiation field and the
X-ray emission have to be well known to use the UV lines as mass-loss
rate diagnostics.  There are several factors that can influence
mass-loss rate determinations, and it is worth noting these and
their relative importance as briefly discussed below.

Although the stellar radiation field is well described
by our models (see Fig.\,\ref{fig:xiop}), the detailed photospheric
models \citep[e.g.][]{mb2008, mor2008}) give temperatures with
uncertainties of several thousand Kelvin.  Our models show that
such differences in temperature are sufficient to alter the ionization
stratification significantly.  Thus, uncertainties in $T_{\rm eff}$
are a source of uncertainty in \mdot.

For our models  we adopted abundances as derived by
\citet{mor2008}. The uncertainty in abundances would reflect on the
mass-loss rate determinations. However, we determine mass-loss rates only
within a factor of few, which is a higher uncertainty than
could be explained by the abundances.

Another source of uncertainty is represented by the spectral and spatial
specifics of the X-ray radiation. From observations we constrain the
X-ray flux quite well; however, the X-ray spectral distribution is only
roughly known for our sources, especially for the X-ray fainter stars
(see Table\,\ref{tab:xspecmod}).  In addition, the location of the X-ray
plasma throughout the wind is unclear.

Similar to other non-LTE stellar atmosphere models, the PoWR model
assumes that the statistical equilibrium is established locally. In a
situation when the densities and the wind flow times are small, the
timescale for recombinations can become longer than the timescale for
transport by advection. In their studies of low density winds,
\citet{mar2005} tested the effects of advection and adiabatic cooling on
the ionization structure and the UV lines for a star with
$\log{\dot{M}}\approx -9$. They found that the inclusion of these
effects would result in an increase of the empirically derived log $\dot{M}$ 
by $\sim$0.15. This is within the error margin of the \mdot\ derived
for our sample stars. It is also important to notice that our final
results are based on unclumped wind models. If the winds are clumped,
the recombination time scale in dense clumps becomes shorter. As a test
we calculated a grid of models that account for microlumping for \bcep.
However, no significant difference was found compared to the un-clumped
models. Moreover, in our models, the N\,{\sc v} and C\,{\sc v} produced
via Auger ionization are the leading ionization stages, therefore our
assumption of local ionization balance does not introduce a large error
in the model ionization structure.

We assume a spherically symmetric winds. Except \tsco\ that
has a complex magnetic topology, the stars in our sample are oblique
magnetic rotators. Their winds are likely confined by magnetic
fields, therefore, the treatment by the spherically symmetric models
can be a source of uncertainty.

Four of our program stars are $\beta$\,Cep type stars.  As
noticed by \citet{don2001}, in principle, the mass-loss rate may
change by $\approx 20$\%\ in response to the $\approx 9$\%\
pulsation-induced luminosity changes, while the terminal velocity
should remain roughly constant.  However, these changes in the
\mdot\ are within the errors of our estimates.

Ultimately it appears that better determinations of the effective
photospheric temperature along with a better understanding of the
flow structure and X-ray sources in the systems will be needed to
obtain consistent mass-loss rates that result in self-consistent
fits for all of the UV wind lines.

\subsection{The weak wind problem}
\label{sec:ww}

It is known that O-type dwarfs with luminosities $\log{L/L_\odot}\lsim
5.2$ have mass-loss rates that are orders of magnitude lower than
predicted by the CAK theory (see \citet{marcolino2009} and references
therein). This discrepancy is commonly refereed to as ``the weak-wind
problem''.

\citet{bab1996} refined the CAK theory for the B-type stars by
studying the theoretical effects of shadowing by photospheric lines
on radiative acceleration. It was shown that these effects have
large consequences on the winds of main sequence B~stars. In particular,
the main difference from the predictions of the CAK theory is found
for stars with $T_{\rm eff}\approx 20000\,-\,23000$\,K and
$\log{g}\approx3.7\,-\,4.0$. For these objects, the mass-loss
rate is found to be lower than predicted by CAK theory by at least
a factor of 4. Among the stars in our sample, \zcas\ has values of
\Teff\ and $\log{g}$ for which the theoretical wind models were
computed by \citet{bab1996}. The predicted mass-loss rate for this
star is $10^{-8.8}$\,\myr with a terminal wind velocity of $v_\infty
= 3.4v_{\rm esc}$ (compared to $\log{\mdot} = -8.2, v_\infty =
1.8v_{\rm esc} $ as predicted by the CAK theory).

Our results show that the mass-loss rate in \zcas\ does not exceed
$10^{-9.7}$\,\myr. This is $\approx 8$ times smaller that the
prediction of \citet{bab1996}. Furthermore, we do not find any
evidence for the predicted fast wind velocity of $v_{\infty} =
2100$\,km\,s$^{-1}$.  Similarly, for all other stars in our sample,
the mass-loss rates are an order of magnitude lower than predictions
by \citet{abb1982} based on the CAK theory. The discrepancy with the
predictions of \citet{vink2000} is even larger.  Thus we conclude that
all stars in our sample belong to the category of weak-wind
stars. \footnote{Recently \citet{lucy2010} identified a weak-wind
  domain on a $\log{g}$--$\log{T_{\rm eff}}$ diagram where a star's
  rate of mass loss by a radiatively-driven wind is less than that due
  to nuclear burning. All our program stars belong to this domain.}

This conclusion is corroborated by the high {\em work ratio $Q$}
we find in our program stars (see Table\,\ref{tab:stellarparameters}).
As the work ratio $Q$ we define the mechanical work per unit time done
by the radiation field compared to the mechanical luminosity of the
wind., The exact definition of $Q$ as computed in the PoWR code is
\begin{equation}
Q\equiv \frac{\int_r \left[g_{\rm rad}(r) - \frac{1}{\rho(r)}
\frac{{\rm d}P_{\rm g}}{{\rm d}r}\right]{\rm d}r}
{\int_r \left[v\frac{{\rm d}v}{{\rm d}r} +
\frac{GM_\ast}{r^2}\right]{\rm d}r},
\label{eq:q}
\end{equation}
where $g_{\rm rad}$ is the radiative acceleration, $P_{\rm g}$ is the gas
pressure, and other symbols have their usual meanings.  In a hydrodynamically
consistent model $Q=1$. When $Q>1$ the model predicts that there is
sufficient line opacity  to produce radiative acceleration that is
capable to drive stronger wind than the observed from the  UV resonant
doublets. We calculated hydrodynamically consistent models by choosing
\mdot\ values that that yield work ratios of $Q\approx 1$. Such models
yield $\log{\mdot(Q\approx 1)}=-9.4$ for \voph\ and $\log{\mdot(Q\approx
1)}=-8.2$ for \xicma. 

Importantly, these work ratios $Q$ are calculated from models that
include X-ray emission in the ionization balance. \citet{drew1994}
highlighted the effect that X-ray emission may have on the wind velocity
and mass-loss rate. They suggest that ionization by X-rays may change
the ionization structure in the inner part of the wind, thus reducing
the total radiative acceleration and consequently the mass-loss rate.
Our models include the ionization by X-rays in the wind acceleration
zone at the level and temperatures indicated by the X-ray observations.
Yet, we find that there is still sufficient radiative acceleration to
drive mass-loss  in excess of the value inferred from the UV
line-fitting analysis. Therefore, the ionization by X-rays cannot be the
unique solution of the weak-wind problem.

\citet{cas1994}, \citet{cass1994}, \citet{coh1997} studied the X-ray
emission from near-main sequence B-type stars based on {\em Rosat} data.
They found that these stars show departure from the $\Lx\approx
10^{-7}\Lbol$ law which holds for O-type stars. From evaluating the
emission measure of X-ray emitting gas, they concluded that a major
fraction of the wind emission measure is hot, whereas in shocked wind
theory less than 10 percent of the wind emission measure should be hot.

Our observations and models support these conclusions. The models show
that the Emission Measure 'filling factor' of the hot material (as
determined from X-ray observations) relative to the cool wind $X_{\rm
  fill}$ exceed unity for all our stars. While $X_{\rm fill}\approx 8$
for \tsco\ is smallest, it is as large as $X_{\rm fill}\approx 200$
for \zcas\ and extremely large $X_{\rm fill}\approx 1000$ for
\xicma. Thus, not only are the UV lines in \xicma\ the most difficult
to reproduce from the models, its X-ray filling factor and wind work
ratio $Q$ are also outstandingly high.

The high X-ray filling factors imply that either the hot gas $EM_{\rm
  hot}$ occupies a much larger volume than the cool wind $EM_{\rm
  cool}$ where the UV lines are formed, or that the X-ray emitting gas
has higher density.  Both possibilities seem plausible in the case of
magnetic stars. The cool wind can occupy a relatively smaller volume
because it emerges primarily from the magnetic polar regions in stars
with dipole fields. On the other hand, high density hot plasma can
occur in magnetically confined wind zones or loops.

\section{Discussion}
\label{sec:disc}

\subsection{X-ray emission from magnetic early-type B stars}

{\changed One of the results of our study is that X-ray properties of
  newly discovered magnetic B-type stars are diverse.  Except for
  \tsco\ and \xicma, our program stars are not especially X-ray
  luminous.  On the contrary, \voph\ and \zcas\ are intrinsically
  X-ray faint. Can these differences in luminosity be explained in the
  framework of the MCWS model? \citet{bma1997} obtained a simple
  scaling relation between X-ray luminosity and the magnetic and
  stellar wind parameters in the framework of their MCWS model. They
  predicted that the X-ray luminosity scales approximately with the
  product of wind momentum and some power of the magnetic field
  strength, $L_{\rm X}\propto B^{0.4}\dot{M}v_\infty$ (see their
  Eqs.\,(10)--(11)).  We apply this scaling relation to predict X-ray
  luminosities for our sample stars with dipole magnetic fields. Among
  our sample stars, \xicma\ has the strongest magnetic field, while
  its wind momentum is comparable to other stars. The scaling relation
  of \citet{bma1997} predicts rather weak dependence on the magnetic
  field strength. The predicted value of
  \Lx\ ($4\times10^{30}$\,erg\,s$^{-1}$) is an order of magnitude
  lower that the observed.  On the other hand, \voph\ and \zcas\ have
  a low wind kinetic energy and a weak magnetic field. Using the lower
  values of $\dot{M}$ from Table\,\ref{tab:stellarparameters}, the
  predicted X-ray luminosities for these stars ($\approx {\rm few}
  \times10^{29}$\,erg\,s$^{-1}$) agrees well with the observed ones.
  Similarly, for the lower value of $\dot{M}$ for \bcep, the predicted
  and observed luminosities agree quite well. Thus, the basic scalings
  obtained by \citet{bma1997} for the X-ray luminosity can
  qualitatively explain the difference in the level of X-ray
  luminosity among \bcep, \voph, and \zcas. Note, however, that using
  the upper values for \mdot\ for these stars in the \citet{bma1997}
  scaling relation, would result in the X-ray luminosities that are
  too high compared to the observed ones.

The more serious challenge is to understand the comparatively low
temperatures of X-ray emitting plasma obtained from the analyses of
the observed spectra and the lack of time modulations of the X-ray
flux.} The X-ray spectral temperatures of magnetic B-stars in our
sample are not especially high.  \zcas\ has one of the softest X-ray
spectra among those measured in OB stars. Moreover, X-ray emissions
from both \tsco\ and \bcep\ were monitored throughout a rotation
period, and neither display rotational modulations in their X-ray
light curves \citep{fav2009, ign2010}.

Compared to non-magnetic stars, {\changed the harder X-ray
  spectrum} and rotationally modulated X-ray variability are
predicted by {\em different} models of hot plasma production in
magnetically confined winds \citep[e.g.][]{li2008,gag2005}.  Can these
models, which assume that wind motion is governed by the magnetic
field, be applied to our program stars?

It is usual to describe the relative importance of magnetic fields
in gases by the plasma-$\beta$ parameter with
$\beta_{\rm p} =8\pi p/B^2$, where $p$ is the
gas pressure.  The gas is magnetically dominated when
$\beta_{\rm p}<1$.

For supersonic flows such as stellar winds the ram pressure exceeds
the gas pressure and the dynamical importance of a magnetic field is
defined by the ratio of wind kinetic to magnetic energy density, given
by the supersonic flow $\beta$ viz. $4\pi\rho v^2/B^2$, where $v$ is
the supersonic flow speed. A small or large value indicates whether
the magnetic field locally dominates the bulk motion, or vice
versa. Supersonic rotation can be treated similarly (e.g. Brown et al
2008). Specifically \citet[][and ref.\ therein]{alt1969} show that
beyond the {\em Alfv{\'e}n} radius $R_{\rm A}$, where

\begin{equation}
B^2/8\pi=\rho v^2/2
\label{eq:rw}
\end{equation}
is met, the radial stellar wind forces magnetic field lines to
to become approximately radial. By contrast for $R<R_{\rm A}$, the
wind flow is confined by the magnetic field.

In the context of
stellar winds, wind confinement was considered by Babel \&
Montmerle (1997), ud-Doula \& Owocki (2002), Brown et al (2008) and others. 
Recalling that the wind density
$\rho=\dot{M}/4\pi v(R) R^2$, one can express $R_{\rm A}$ from
the condition set by Eq.\,(\ref{eq:rw}):
\begin{equation}
R_{\rm A}=\sqrt{\frac{\dot{M}v(R)}{B^2}}.
\label{eq:RA}
\end{equation}
In case of a dipole magnetic field $B=B_0(R_\ast/R)^3$  Eq.\,(\ref{eq:RA})
can be re-written as
\begin{equation}
r_{\rm A}\cdot w(r)^{\frac{1}{4}}= \left(\frac{R_\ast^2B^2}
{\dot{M}\vinf}\right)^{\frac{1}{4}}
\equiv \eta_\ast^{\frac{1}{4}},
\label{eq:betastar}
\end{equation}
where $r=R/R_\ast$, and the standard parametric wind velocity law with
$v(r)=v_{\infty}(1-r^{-1})^\beta\equiv v_{\infty}w(r)$ is assumed.
It is convenient to express $\eta_\ast$ in terms of normalized stellar
wind parameters, with
$\dot{M}_{-9}$ the mass loss rate in units of $10^{-9}$\myr, \vinf\ in 
km\,s$^{-1}$, and $\cal{R}_\ast$ the radius in units of solar radius
$R_\odot$. Then
\begin{equation}
\eta_\ast\approx 1 \cdot \frac{{\cal R}_{\ast}^2} {\dot{M}_{-9}
v_\infty}\cdot {B_0^2}.
\label{lsu}
\end{equation}

For stars with dipole fields (i.e.\ excluding \tsco) the field
strength at the magnetic pole is shown  in Table\,\ref{tab:stellarparameters}.
To estimate $r_{\rm A}$ we use as a characteristic value the field
strength at the magnetic equator (i.e.\ half of the polar field
strength).   Then we find
$\eta_\ast(\bcep)\msim 3\times 10^3$,
$\eta_\ast(\xicma)\approx 6\times 10^6$,
$\eta_\ast(\voph)\msim 4\times 10^3$, and
$\eta_\ast(\zcas)\msim 6\times 10^3$.
Consequently, the Alfv\'en radius is at $\approx 50$\,\Rstar\  for
\xicma\ and $\lsim 10$\,\Rstar\ for other stars.  These estimates  are
based on the highest mass-loss rates obtained from fitting of the UV
lines.

However, we know that a significant fraction of the wind mass
loss can be in the form of hot X-ray emitting plasma. We may hypothesize
that the ``true'' total amount of matter is much higher. We calculated a
hydrodynamically consistent model for \voph\ and \xicma\ (see Section
\ref{sec:ww}) to obtain a mass-loss rate $\log{\mdot(Q\approx 1)}=-9.4$
for the former and $\log{\mdot(Q\approx 1)}=-8.2$ for the latter. Even
with these higher mass-loss rates, the magnetic fields regulate
the wind motion up to $\approx 7$\,$R_\ast$ for \voph\ and $\approx
20\,R_\ast$ for \xicma.

Thus the winds of our sample stars are strongly confined out to
several stellar radii where the wind velocity should reach its
terminal value.  Due to the very weak signatures of the winds seen
in the UV line profiles, we can establish \vinf\ only with a limited
degree of accuracy. Using $v_\infty\approx 700$\,km\,s$^{-1}$ is
reasonably consistent with the UV lines.  Colliding wind streams
at such a speed from two opposite directions of the magnetosphere
should produce strong shocks, heating plasma up to $\approx 20$\,MK.
This is far above the maximum temperatures we infer from the spectral
analysis (see Table\,\ref{tab:xspecmod}).

From the UV line modeling, we cannot exclude somewhat lower \vinf\
for the winds of our stars, which would alleviate the expectation
of very hot temperatures, scaling as $v_\infty^2$.  However, there
is, perhaps, a more severe problem of the differential emission
measure distribution (DEM).  The emission measure of the hottest
plasma is $\approx 35$\%\ of the total EM in \xicma\ and only
$\approx$10\%\ in \bcep. On the other hand, the plasma with the
largest emission measure has a temperature of only about 1.3\,MK
in \xicma\ (see Table\,\ref{tab:xspecmod}). This dominant contribution
to the DEM from the cooler hot gas component is in apparent
contradiction with the predictions of the MCWS models, where the
DEM is expected to peak at the hottest temperature \citep[e.g. see
Fig.\,8 in][]{naze2010}.

What could be the source of wind heating in magnetic $\beta$\,Cep-type
stars?  As pointed out by \citet{fav2009}, the relatively low
temperature of the X-ray emitting plasma and the absence of flares
makes heating due to magnetic reconnection in analogy to active
cool stars rather implausible.  Heating by wind shocks owing to the
intrinsic LDI mechanism in radiatively driven winds is a possibility.
However, the LDI needs further theoretical study in relation to its
operation in complex geometries and the dynamics of magnetically
confined winds.

Another possibility could be heating by the deposition of mechanical
energy related to stellar pulsations. \citet{cas1996} found from EUVE
observations that the $\beta$\,Cep-type star $\beta$\,CMa has an EUV
exceeding atmosphere model predictions by an order of
magnitude. \footnote{It is possible that this problem can be resolved
  by a better description of X-ray emission in model atmospheres
  (Hamann \etal, in prep.).}. They suggest that this may be related to
the presence of heated regions near the stellar surface owing to
pulsations and magnetic fields. If the upper atmosphere is heated
mechanically by pulsational effects, it is likely that the X-ray
source regions are also heated by a related mechanism. In this case,
the dynamic time scale of the wind is smaller than the time required
to establish ionization equilibrium.

\begin{figure}
\centering
\includegraphics[width=0.9\columnwidth]{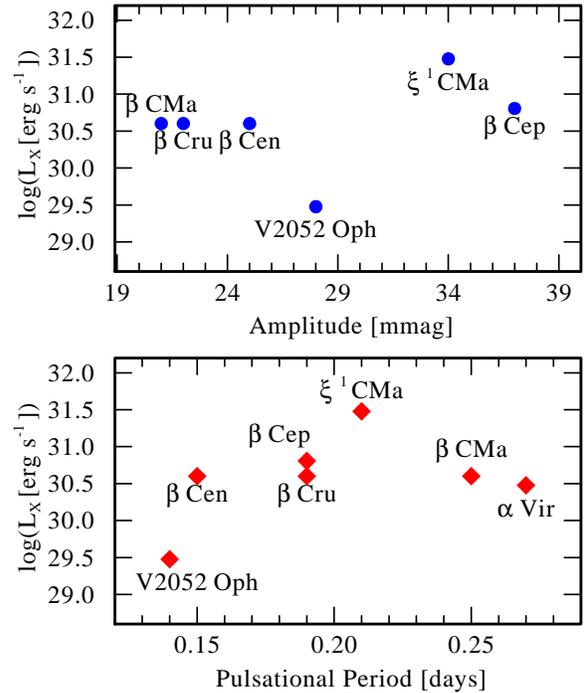}
\caption{\changed X-ray luminosity of some prominent $\beta$\,Cep-type 
stars vs. magnitude of pulsation (upper panel) and pulsational period 
(lower panel) from the catalog of \citet{st2005}.}
\label{fig:ppuls}
\end{figure}

{\changed To investigate this possibility, we show in
  Fig.\,\ref{fig:ppuls} the X-ray luminosities versus pulsation
  periods and amplitudes for few prominent
  $\beta$\,Cep-type variables. Given the small sample it is difficult
  to draw any conclusions, yet there is no apparent dependence of the
  X-ray luminosity on period or pulsational amplitude.}

\subsection{Comparison of X-ray properties between B stars with and
without confirmed magnetic fields}

In this section we compare X-ray properties of ``normal'' early-type B
stars where magnetic fields have not been detected up to now. The
\xmm\ spectrum of the $\beta$\,Cep-type variable $\beta$\,Cen (B1III)
is well fitted with a three temperature spectral model, with the
hottest component $kT_{\rm max}\approx 0.6$\,keV and $\left< kT
\right>=0.3$\,keV \citep{raas2005}.  \citet{hub2006} and
\citet{sil2009} searched for a magnetic field in the $\beta$\,Cep-type
$\beta$\,CMa (B1II-III) and did not detect magnetism, with errors on a
longitudal field of just 10\,G.  $\beta$\,CMa was observed by \xmm,
its spectrum is well described by a multi-temperature model with the
hottest temperature component at $kT=0.3$\,keV (Waldron \etal\ 2011,
in prep.).  {\em Chandra} LETGS spectrum of $\alpha$\,Vir (B1III-IV)
was analyzed in \citet{zp2007}, who report the maximum of the emission
measure distribution as being between 0.2\,keV and 0.3\,keV.  We
obtained LETGS {\em Chandra} observations of $\alpha$\,Cru
(B0.5IV). Its X-ray temperature and luminosity are similar to those of
$\alpha$\,Vir (Ignace \etal\ in prep.).  $\beta$\,Cru is another
B1~giant observed by {\em Chandra}.  \citet{zp2007} find the maximum
of the emission measure distribution in this $\beta$\,Cep-type
variable is between 0.2\,keV and 0.3\,keV.  Thus, it appears that
X-ray luminosities and temperatures are very similar among the
\bcep-type stars.

On the other hand, the X-ray luminosities of magnetic stars have much
greater dispersion, with two significant outliers: the X-ray luminous
\xicma\ and the X-ray faint \voph, while their spectral X-ray
temperatures are rather similar (see Table\,\ref{tab:xspecmod}).
{\changed Due to the small number of stars included in our
  investigation, this conclusion is only preliminary. Larger samples
  need to be considered to draw firm conclusions about the differences
  in X-ray properties of magnetic and ``normal'' early-type B stars.

Overall, it has emerged from our analysis that high X-ray luminosity
and hard X-ray spectra are not necessary observational characteristics
of the magnetic early-type stars. This conclusion fully confirms the
earlier results based on the {\rm RASS} data (see also recent results
on X-rays from A0p stars \citep{rob2011}).

One of the complications in massive early-type star studies is that a
significant fraction of such stars is in binary or multiple systems
and may have spatially unresolved companions that are intrinsic X-ray
sources.  An X-ray source spatially coincident with a B star can be
due to a lower mass companion \citep{cz2007, evans2011}. In this paper
we consider B0-B2 type stars. These stars are young, not older that a
few 10\,Myr.  Possible low-mass companions would be coronally active
with X-ray luminosities of $10^{28}$--$10^{30}$\,erg\,s$^{-1}$
\citep[e.g.][]{flac2003}. Two newly X-ray detected stars considered in
this paper, \voph\ and \zcas, have X-ray luminosities in this range
(see Table\,\ref{tab:xray}). Therefore, in principle, we cannot rule a
confusion with an unseen coronal component. Nevertheless, we have
strong indications that the observed X-ray emission is intrinsic for
the early-type B stars. The spectra of X-ray emission from \zcas\ and
\voph\ are soft, e.g.\ in \zcas\ the average temperature is
1\,MK. This is not usual for young T\,Tau type stars, which have
average temperatures in the range 5--30\,MK \citep{gn2009}. The
temperature of the X-ray emitting plasma in \voph\ is similar to other
\bcep-type stars, albeit its X-ray luminosity is quite low (see
Fig.\,\ref{fig:ppuls}). From the analysis of the UV spectra of our
program stars, we know that X-rays are present in the winds since they
are required to enhance the fraction of C\,{\sc iv} and N\,{\sc v} in
the winds (see Sect.\,\ref{sec:voph}). On this basis, we believe that
the observed X-ray emission is intrinsic for \bcep, \voph, \zcas, and
\xicma. Among stars in our sample, \bcep\ is a known binary star with
a late Be-type companion. Late Be stars are not known to be intrinsic
X-ray sources, therefore it seems reasonable to assume that the
observed X-ray emission is intrinsic to the B2 component.}

\subsection{X-rays in stars with very small \mdot}

\zcas\ displays quite a soft X-ray spectrum with nearly 85\%\ of the EM
at temperature $\lsim 1$\,MK (Fig.\,\ref{fig:zcassp} and
Table\,\ref{tab:xspecmod}).  Our estimate of its mass-loss rate
($\mdot\sim 10^{-11}$\,\myr) places this object in the regime where the
winds may display unusual properties.

\citet{sp1992} demonstrated that the assumption of a one-component fluid
will not be valid for low density winds. Instead, the metal ions lose
their dynamical coupling to the ions of hydrogen and helium.  The metal
ions will move with high velocities, with helium and hydrogen failing to
be dragged along. The collisionally induced momentum transfer is
accompanied by frictional heating, which dominates the energy balance.
As a result, \citet{sp1992} predict electron temperatures of $T\sim
1$\,MK in the outer wind regions for stars with mass-loss rates of
$\approx 10^{-8...9}$\,\myr. The subsequent analysis of \citet{kk2000}
and \citet{op2002} favored even weaker winds with $\mdot\sim
10^{-11}$\myr\ for the wind decoupling to occur.

The theoretical analysis of this problem is not settled
\citep[e.g.][]{vor2007}. \citet{go1994} investigated the effects of
``Doppler heating'' in stars with low mass-loss rates, and confirmed
the conclusions of \citet{sp1992}. They also noticed that Doppler
heating may lead to an instability, possibly resulting in a larger
degree of microturbulence. This is an interesting prediction since our
analysis favors large microturbulent broadening, as in the case of
\xicma.

Given its small \mdot\ value and very soft X-ray spectrum, it is
tempting to suggest that \zcas\ may be a candidate object for wind
decoupling.

Another interesting mechanism for generating X-ray emission in
rotating magnetic stars was put forward by \citet{suz2006}. They
proposed that extended X-ray-emitting regions can exist in massive
stars with thin winds owing to collisionless damping of fast MHD
waves. Stellar rotation causes magnetic field lines anchored at the
stellar surface to form a spiral pattern, and magneto-rotational
winds can be driven (e.g., Weber \& Davis 1967). If the structure
is magnetically dominated, fast MHD waves generated at the surface
can propagate almost radially outward and cross the field lines.
The propagating waves undergo collisionless damping owing to
interactions with particles surfing on magnetic mirrors that are
formed by the waves themselves. The dissipation of the wave energy
produces heating and acceleration of the outflow. The \citet{suz2006}
mechanism works effectively in moderate and fast rotating stars.
Considering Figure\,6 in \citet{suz2006}, the magnetic field strength,
wind density, and rotation speed in \zcas\ are such that the star
may be in the domain where the Suzuki \etal\ mechanism is operational.

\subsection{On the multi-phase structure of winds from magnetic
early B-type stars}

We try to understand whether some of the facts that have emerged from our
analysis can be qualitatively explained by a picture of an oblique
magnetic rotator where the bulk of the stellar wind is governed by the
dipole magnetic field.

Despite relatively strong surface fields, oblique magnetic rotators
have open field regions near their magnetic poles, and thus may
have sectors of wind flow that are similar to normal B~type stars.
The UV spectra could originate in these cool sectors of the wind
with a smaller filling factor as compared to the hot gas. The wind
distribution is clearly asymmetric (not even axisymmetric owing to
rotation).  Perhaps this could explain the strange properties of
the UV lines seen in \xicma\ that we view nearly rotational pole-on
to the rotation axis, and possibly the large dispersion in X-ray
luminosities among our sample stars as well.

The hot part of the wind is apparently heated up to a few MK. This matter
is confined by the magnetic field up to $10\,\Rstar$, so it occupies a
large volume, and, in parts, may have high density. This would explain
the large filling factors of hot material we infer from our models. The
spatial separation between hot and cool gas may explain  how
the observed N\,{\sc v} doublet can be weaker than predicted by our
stellar atmosphere models. The total amount of matter in the cool and
hot wind components may be close to values predicted by our
hydrodynamically consistent models, and so the ``true'' mass-loss rates
are a factor of a few higher than obtained from the empirical fits of UV
lines using a model that relies on spherical symmetry.



\section{Summary of results}
\label{sec:con}

Dedicated \xmm\ observations were obtained for three early-type
magnetic B-type stars, \xicma, \voph, and \zcas. We report first
detections of X-ray emission from \voph\ and \zcas. The observations
show that the low-resolution X-ray spectra of our program stars are
well described by a multi-temperature CIE plasma. The bulk of the
emission measure originates from the plasma with a temperature of
$\approx 1$\,MK.  The X-ray luminosities differ by large factors, with
\xicma\ being the most X-ray luminous star in our sample, and
\voph\ the least luminous.

\medskip\noindent We compare X-ray properties of $\beta$\,Cep-type
variables that have magnetic field detections against those without such
detections. Our comparison shows {\changed indications} that the general
X-ray properties are quite similar, although the magnetic stars display
a greater dispersion in their X-ray luminosities. We searched for
correlations between X-ray emission and pulsational and rotational
properties of the stars but were not able to find any.  Larger samples
need to be studied to draw firm conclusions. 

\medskip\noindent  The basic MCWS model scaling relation between X-ray
  luminosity and stellar magnetic field and wind parameters
  \citep{bma1997} qualitatively describes the difference in the observed
  level of X-ray emission from our sample stars. 

\medskip\noindent
We compiled a sample of peculiar early B-type magnetic stars and
considered their X-ray properties based on archival data. The
comparison shows that these stars, despite very similar stellar and
magnetic field parameters, have quite different X-ray properties:
while some stars are hard and luminous X-ray emitters,
others are apparently soft and rather faint. {\changed This new data
confirm earlier results based on studies of Ap-Bp stars using RASS
data.}

\medskip\noindent
We analyzed spectra of five non-supergiant B~stars with magnetic fields
by means of non-LTE comprehensive stellar atmosphere code PoWR. The PoWR
models accurately fit the stellar photospheric spectra and
also accurately reproduce the SED from the UV to the IR band.

\medskip\noindent
The PoWR code was used to empirically obtain wind parameters
of \tsco, \bcep, \xicma, \voph, and \zcas\ from the analysis of
C\,{\sc iv}, N\,{\sc v}, and Si\,{\sc iv} doublets in stellar UV
spectra.

\medskip\noindent The model UV lines depend sensitively on the
parameters of X-ray emission, wind velocity, and mass-loss rate.
X-ray properties as derived from the observed spectra were
incorporated into the wind models.  We confirm that the emission
measure filling factors of X-ray emitting material are high, 
exceeding unity for all stars.

\medskip\noindent Our analysis revealed the weak wind problem for the
magnetic early type B~stars. The wind terminal speeds are on the order
of stellar escape speed. The inferred mass-loss rates from the UV line
analyses are significantly lower than the theoretically expected
and predicted by hydrodynamically consistent wind models.

\medskip\noindent Although X-rays strongly impact the ionization
structure in the wind, their  effect does not reduce the total radiative
acceleration enough to explain the low mass-loss rates deduced from
modeling the UV lines.  When X-rays at the observed level and
temperatures are included in the model, there is still sufficient
radiative acceleration to drive a stronger mass-loss than the observed
one.

\section*{Acknowledgments}  Based on observations obtained with
\xmm, an ESA science mission with instruments and contributions
directly funded by ESA Member States and NASA. {\changed We are grateful
  to Nolan Walborn for his comments on the manuscript and his advice
  on the spectral classification of our sample stars. We are also
  indebted  to the referee, Stephen A. Drake, for very useful comments that
  helped to improve the clarity of the paper, and for pointing out some
  aspects in the studies of magnetic stars that we did not consider in
  the original manuscript. } This research has made use of NASA's
Astrophysics Data System Service and the SIMBAD database, operated at
CDS, Strasbourg, France.  Funding for this research has been provided
by NASA grant NNX08AW84G (RI), DLR grant 50\,OR\,0804 (LMO) and a UK
STFC Grant (JCB).

\bibliographystyle{mn2e}
\bibliography{magbs.bib}

\label{lastpage}
\end{document}